\documentclass{article}

\usepackage{arxiv}

\usepackage[utf8]{inputenc} % allow utf-8 input
\usepackage[T1]{fontenc}    % use 8-bit T1 fonts    % hyperlinks
\usepackage{url}            % simple URL typesetting
\usepackage{booktabs}       % professional-quality tables
\usepackage{amsfonts}       % blackboard math symbols
\usepackage{nicefrac}       % compact symbols for 1/2, etc.
\usepackage{microtype}      % microtypography
\usepackage{lipsum}		% Can be removed after putting your text content
\usepackage{graphicx}
\usepackage{natbib}
\usepackage{doi}
\usepackage{newtxtext}
\usepackage{newtxmath}
\usepackage{authblk}
\usepackage{tabu}
\usepackage{multirow}

\title{Energy Transfer Mechanisms in Wake-Modulated Transonic Flutter}

\date{} 					% Or removing it
\author[1]{Vedasri Godavarthi}
\author[2]{Jacob Turner}
\author[1]{Jung-Hee Seo}
\author[1]{Rajat Mittal}%\thanks{mittal@jhu.edu}
\affil[1]{Department of Mechanical  Engineering, Johns Hopkins University, Baltimore, MD 21210, USA}
\affil[2]{Department of Mechanical Engineering, Colorado State University, Fort Collins, CO 80523, USA}

\hypersetup{
pdftitle={A template for the arxiv style},
pdfsubject={q-bio.NC, q-bio.QM},
pdfauthor={David S.~Hippocampus, Elias D.~Striatum},
pdfkeywords={First keyword, Second keyword, More},
}

\begin{document}
\maketitle

\begin{abstract}
	Transonic flutter is a detrimental aeroelastic instability that can generate large-amplitude structural oscillations, leading to severe vibration, fatigue damage, reduced operational limits, and potentially catastrophic structural failure. Incoming wake disturbances can further amplify this instability, making it critical to identify the underlying aerodynamic mechanisms responsible for predicting and controlling flutter onset. The underlying flow physics is complex with nonlinear interactions between the wake and the wing, shock motion, shock-induced flow separation, vortex shedding and the wing motion. In this study, we perform high-fidelity direct numerical simulations of a sinusoidally pitching NACA0012 airfoil with an underwing cylinder at various transonic Mach numbers and a Reynolds number of 10,000. Through energy maps, we identify that the addition of the cylinder significantly expands flutter boundaries compared to an airfoil-only system. We extend the force partitioning method to partition the power transferred between the flow and the airfoil for compressible flows. Application of this approach to distinct regions of the flow  domain indicates that the gap flow between the wing and the cylinder is the dominant contributor to the energy transfer from flow to the wing. The blockage effects due to the cylinder cause flow acceleration on the wing which further enhances the tendency for flutter. We investigate cylinder placement relative to the airfoil to reveal that flutter is enhanced only when the cylinder is placed upstream of the pivot point on the airfoil. The current study highlights how such partitioning methods can parse force and energy transfer mechanisms in complex, unsteady high-speed flows.
\end{abstract}

% keywords can be removed
%\keywords{First keyword \and Second keyword \and More}

	\section{Introduction}
	\label{sec:intro}

Transonic flutter is a challenging aeroelastic problem due to the nonlinear interaction between the structural dynamics and the compressibility effects in transonic regimes. Several studies have identified a transonic dip phenomenon wherein the flutter onset speed is reduced in the transonic flow regime \citep{Lee2001,Schewe2003,Kholodar2004,Dietz2006,Bendiksen2011} thereby limiting the operational regime of commercial and military aircraft. The flutter instability causes growth of oscillations but often saturates into detrimental limit-cycle oscillations due to the nonlinearities rather than diverging monotonically \citep{Schewe2003,Lambourne1964,Dietz2006}. These limit cycle oscillations can cause structural damage of the aircraft through fatigue loading and can also cause pilot discomfort in high-performance aircraft. The dominant nonlinear behaviour is attributed to the coupling between unsteady shocks and separated flow, including $\lambda$-shock formation, shock-induced stall and viscous effects in the boundary layer \citep{GhoshChoudhuriKnight1996,Yamasaki2004,ThomasDowellHall2004,Karnick2017,CorkeThomas2015,Turner2024}. Hence, determining flutter boundaries and understanding the relevant flow mechanisms is essential for expanding the operational regimes of aircraft . 

While transonic flutter already involves highly complex flow physics in nominally undisturbed freestream conditions, practical flight environments often include incident flows modified by upstream wakes. Such wake–wing interactions are particularly relevant in formation flight, where the aerodynamic environment experienced by the follower aircraft depends strongly on its position within the wake of the upstream wing \citep{NingKrooAftosmisNemecKless2014}. In transport aircraft, wakes generated by engine nacelles have also been shown to alter shock–boundary-layer interactions on the lower wing surface, leading to buffet-like low-frequency shock oscillations and associated unsteadiness \citep{SpinnerRudnik2024,LuerkensMeinkeSchroeder2024}. Furthermore, wakes shed from upstream wings undergoing buffet instability can induce significant unsteady aerodynamic loading on downstream tail-plane surfaces \citep{SchauerteSchreyer2024,KleinertStoberLutz2023,KleinertEhrleWaldmannLutz2024}.
    
The effects of wing-wake interactions on flutter instability are extensively observed in high performance aircraft mounted with external underwing payloads such as fuel tank, engines or stores. There have been several studies to understand the flutter instability mechanism through flight tests \citep{Denegri2005,DenegriSharmaNorthington2016} and computational studies \citep{BuntonDenegri2000,Beran2004,Parker2007,Iovnovich2017}. These studies have shown that underwing attachments can also increase susceptibility to flutter and can lower the flutter onset speed. Changes in configuration of these external payloads can substantially modify the aeroelastic response, the onset flutter speed, amplitude of limit cycle oscillations.

The effects of external wakes on wing flutter can be attributed to several mechanisms such as the structural properties of the wing-store configuration, mass distribution, attachment between the store and the wing, transfer of store aerodynamic loads onto the wings and the aerodynamic interaction between the underwing payloads and the wing. The responsible structural mechanisms have been analyzed using complex structural models coupled with inviscid or parametrized aerodynamic models \citep{ThomasDowellHall2002,PadmanabhanPasiliaoDowell2014,PadmanabhanDowellThomasPasiliao2016}. However, the \emph{aerodynamic interaction mechanisms} by which stores modify transonic flutter are yet to be understood. Flight tests have revealed that the addition of the underwing attachments modify the pressure distribution on the lower surface of the wing and can change the local shock pattern in the transonic regime \citep{LeeChen2009,Denegri2013}. Panel-based methods coupled with linearized transonic aerodynamic model suggest that shock interactions associated with the store can strongly influence the response \citep{LeeChen2009,Iovnovich2019}. However, viscous effects can significantly modify shock location, shock strength and the resulting flutter boundary \citep{ThomasDowellHall2004,Karnick2017}.  

Previous studies have incorporated viscous effects through Reynolds averaged Navier--Stokes equations (RANS) and unsteady RANS \citep{ParkerMapleBeran2005,Parker2007, Iovnovich2017,YuanRimplePoirel2021} and shown that viscous effects can substantially modify transonic flutter boundaries relative to Euler-based predictions. However, results from such models are sensitive to the choice of the turbulence model. Further, the fidelity of RANS-based approaches remains limited for isolating the details of the nonlinear aerodynamic interaction mechanisms such as shock-induced separation, vortex shedding. \citet{BeranEtAl2004,Parker2007} conducted Navier--Stokes studies of Goland-type wing configurations. They indicated that the aerodynamic loads of underwing attachments can increase LCO amplitude, while aerodynamic interference beneath the wing can reduce the LCO amplitude. All these studies highlight the sensitivity of aerodynamics to the fidelity of the underlying fluid dynamics solver.
    
For such coupled wing-wake problems, high-fidelity simulations are necessary to fully resolve the vortex-dominated perturbations, wake interactions, gap-flow effects, shock motion and shock-induced separation that may arise between the store and wing in transonic flow. However, a full aircraft configuration with underwing store is extremely complex and challenging. We consider a simplified wake-airfoil system to bridge the gap between a full aircraft configuration and a canonical aeroelastic model in order to obtain a fundamental understanding of the transonic aerodynamics. In transonic flows, rod-airfoil and tandem bluff-body/airfoil configurations have served as canonical models for upstream-wake interaction with a downstream lifting surface \citep{Nakagawa1987,Jacob2005,Boudet2005,Greschner2008}.  While many of these studies were motivated by aeroacoustics or buffet, they establish a computationally tractable configuration in which the fundamental wake-wing interaction can be examined. Motivated by these, we consider a circular cylinder underneath a NACA0012 airfoil in transonic flow conditions as a canonical configuration to enable a fundamental analysis of wake-induced effects on flutter.
\begin{figure}
\centering
    \includegraphics[width=0.8\textwidth]{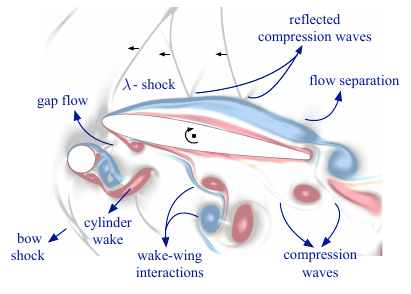}
    \caption{Various flow mechanisms observed in a sinusoidally pitching cylinder-airfoil system in transonic Mach numbers}
    \label{fig:schematic}
\end{figure}

Previously, \citet{Turner2024} used prescribed-motion simulations and energy maps to identify the flutter boundary for a sinusoidally pitching NACA0012 airfoil in transonic flow at a chord Reynolds number of 10,000. They showed that a subcritical instability occurs near $M_\infty\approx 0.7$ and identified shock-induced stall associated with $\lambda$-shock formation as the primary flutter mechanism. The present study extends that framework by introducing a circular cylinder beneath the airfoil as a prototypical model of a wake-generating underwing store. The cylinder--airfoil system is intended to capture, in a minimal setting, the dominant effects of underwing-attachment interference, namely blockage, wake impingement on the wing, modification of the lower-surface shear layer, and the formation of locally supersonic gap flow with multiple compression/shock structures (see figure~\ref{fig:schematic}). We prescribe sinusoidal pitching motion for the cylinder-airfoil system and use energy maps \citep{MenonMittal2019,menon2021initiation} to characterize the flutter boundaries and compare them with those of an airfoil-only configuration. Since the dynamics is observed to be very rich with the presence of shock wave motion, flow separation, vortex shedding and gap flow, we need a quantitative framework to discern the effect of cylinder on wing aerodynamics.

Force partitioning methods have been widely used to identify the flow structures responsible for aerodynamic forces and moments in unsteady incompressible aerodynamic and aeroelastic systems \citep{zhang2015centripetal,menon2021significance,menon2021initiation,prakhar2025modal}. These methods originate from the work of \citet{QuartapelleNapolitano1983} and are closely related to the force-element analysis of \citet{Chang1992}. The underlying framework employs influence potentials to transform surface integrals of pressure forces and moments into distributed volumetric and surface contributions that can be directly associated with specific flow structures and mechanisms. Extensions of these approaches to compressible flows have been relatively limited. \citet{ChangLei1996,ChangSuLei1998} extended the formulation to steady compressible flow by introducing a correction term to the Lamb vector and demonstrated the method on RANS simulations of compressible flow over a delta wing. Related Lamb-vector-based decompositions have also been developed using derivative-moment-transformation integral methods \citep{Wu2007,MeleTognaccini2014,MeleOstieriTognaccini2016}, with additional correction terms introduced to account for transonic compressibility effects.

In the present study, we extend the influence-potential-based force partitioning framework to a moving-body transonic wake-interaction problem in order to diagnose the mechanisms responsible for wake-modified flutter. Since the central focus of this work is the exchange of energy between the flow and the oscillating wing, we further introduce a corresponding ``power partitioning'' framework that decomposes the aerodynamic \emph{power} transferred through the pressure-induced loads into physically interpretable volumetric and surface contributions. This compressible force/power partitioning methodology enables direct identification of the flow structures and regions responsible for destabilizing or stabilizing energy transfer. Although developed here for the transonic cylinder--airfoil configuration, the framework is broadly applicable to high-speed compressible flows involving stationary, moving, and deforming bodies.

The outline of the paper is as follows: Section ~\ref{sec:setup} outlines the numerical setup and computational methodology. Section~\ref{sec:flow_physics} describes the flow physics and energy maps. Section~\ref{sec:cfpm} outlines the compressible force and power partitioning method. We discuss the wake-induced effects in Section 4 and conclusions in Section~\ref{sec:cylinder_influence}. The grid convergence details are presented in Appendix A.

\section{Problem setup}
\label{sec:setup}

We perform high-fidelity, time-accurate, two-dimensional simulations of compressible flow over a NACA0012 airfoil and an airfoil-cylinder system undergoing prescribed sinusoidal motion using high-order sharp-interfaced immersed boundary method (ViCAS3D, Viscous Cartesian All Speed Three-Dimensional [3D] Solver) \citep{turner2024high}. We conduct several simulations at various Mach numbers and pitching amplitudes at a airfoil chord-based ($c$) Reynolds number of $Re_\infty=U_\infty c/\nu_\infty=10,000$, where $U_\infty,\nu_\infty$ are the free-stream velocity and kinematic viscosity, respectively. While this Reynolds number is lower than operational Reynolds numbers for transonic wings, it is sufficiently large so as to generate complex vortex and shock dynamics that are representative of realistic flows, while allowing for a comprehensive simulation-based examination of the large parameter space with a moving boundary. A brief description of the solver and the numerical setup are described below. Further details and a rigorous validation of the solver can be found in \citet{turner2024high}.

 \begin{figure}
\centering
    \includegraphics[width=0.95\textwidth]{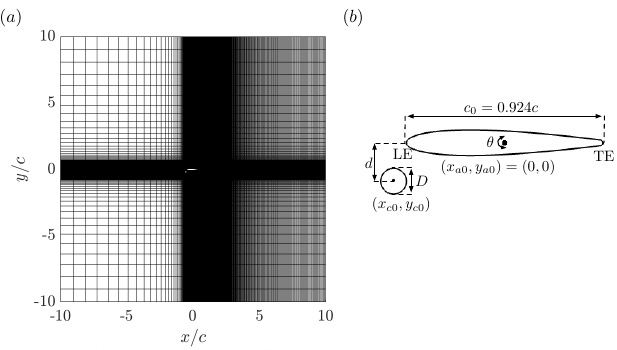}
    \caption{(a) Computational domain and mesh setup for the current study. (b) Setup of cylinder placed underneath the airfoil.}
    \label{fig:domain}
\end{figure}

\subsection{Flow Solver}
We consider two-dimensional compressible Navier--Stokes equations in conservative form:
\begin{equation}
    \frac{\partial \boldsymbol{Q}}{\partial t}+\nabla \cdot \boldsymbol{E} - \frac{M_\infty}{Re_\infty}\nabla \cdot 
    \boldsymbol{F}=0.
    \label{eq:NS_comp}
\end{equation}
\noindent Here, free-stream Mach number is $M_\infty=U_\infty/a_\infty$ where $U_\infty,\,a_\infty$ are the free-stream velocity and the speed of sound, respectively. The free-stream Reynolds number is $Re_\infty=\rho_\infty U_\infty L_c/\mu_\infty$, where $\rho_\infty,\mu_\infty$ are the free-stream density and viscosity and $L_c$ is the characteristic length scale. The vectors $\boldsymbol{Q},\boldsymbol{E},\boldsymbol{F}$ are given by,
\begin{subequations}\label{eq:NS_comp_i}
\begin{align}
\boldsymbol{Q} &= [\rho,\rho \boldsymbol{u},\rho e_t]^T, \label{eq:NS_comp_a}\\
\boldsymbol{E} &= [\rho \boldsymbol{u},\rho\boldsymbol{u}\otimes
\boldsymbol{u}+p\boldsymbol{I},(\rho e_t+p)\boldsymbol{u}]^T,\label{eq:NS_comp_b}\\
\boldsymbol{F} &= [0, \boldsymbol{\tau},\boldsymbol{\tau}\cdot \boldsymbol{u}+\boldsymbol{q}]^T. \label{eq:NS_comp_c}
\end{align}
\end{subequations}
\noindent Here, $\boldsymbol{u}=[u,v]^T$ is the velocity vector in two-dimensions, $p$ is pressure, $\rho$ is density, $e$ is the total energy defined as $e_t=\frac{p}{\left[(\gamma-1)\rho\right]}+\frac{\boldsymbol{u}\cdot \boldsymbol{u}}{2}$ with $\gamma=1.4$ for air and $I$ is the identity matrix. The stress tensor $\boldsymbol{\tau}$ and the heat flux vector $\boldsymbol{q}$ are defined as
\begin{subequations}\label{eq:NS_tau_q_i}
\begin{align}
\boldsymbol{\tau} &= \nabla\boldsymbol{u}+\nabla\boldsymbol{u^T}-\frac{2}{3}(\nabla \cdot\boldsymbol{u})\boldsymbol{I}, \label{eq:NS_tau_q_a}\\
\boldsymbol{q} &=-\frac{\mu}{(\gamma-1)Pr}\nabla T\label{eq:NS_tau_q_b}
\end{align}
\end{subequations}
\noindent where Prandtl number $Pr=0.72$ and $\mu$ is the local dynamic viscosity determined using Sutherland's law \citep{white2006viscous}. The results presented in this work are non-dimensionalized by the respective characteristic length, velocity, time-scale and density as airfoil chord $c$, free-stream velocity $U_\infty$, $c/U_\infty$ and $\rho_\infty$. 

The governing equations are discretized using sixth-order cell-centered compact finite difference schemes, with third and fourth-order compact schemes applied at the boundaries \citep{lele1992compact}. In the current study, high-order compact filtering is implemented with a filter cut-off of $\alpha_f=0.45$ for numerical stability \citep{gaitonde1999high}. These compact filtering and differencing schemes are implemented using the Thomas algorithm \citep{carnahan1969applied}. The governing equations are discretized on a non-uniform Cartesian grid which requires the following transformation for the evaluation of the derivatives. A fourth stage Runge-Kutta scheme is used for time integration \citep{jameson1985numerical}. Additionally, a hyperviscosity method is used for handling the discontinuities caused by shock waves \citet{kawai2008localized}. This method introduces artificial diffusivity biased towards high-wavenumbers that are unresolved. The coefficients for artificial diffusion are added to the dynamic viscosity, thermal conductivity and bulk viscosity based on the fourth-order derivative of the local strain rate and the grid size, following \citet{kawai2008localized,turner2024high}.

Since we are simulating flow over an airfoil and airfoil-cylinder system, the internal bodies are represented by an unstructured surface mesh immersed in a Cartesian fluid mesh. A no-slip boundary condition is implemented on the internal bodies using a sharp-interface immersed boundary method \citep{seo2011high} using the ghost-cell method. The cells that have at least one neighboring fluid cell are identified as ghost cells and are used to implement the no-slip boundary condition. We use a Taylor polynomial interpolant of $\rm{n}^{th}$ order through weighted least-squares minimization. For the current simulations, we use a third-order polynomial interpolant which considers 25 nearest fluid cell solutions to compute the values at the ghost cells. This formulation is the same as the one used in  earlier studies \citep{seo2011high,turner2024high}. For farfield boundaries, an energy transfer and annihilating (ETA) boundary condition is prescribed to avoid spurious reflections \citep{edgar2003general}. 

\subsection{Problem Setup and Mesh}
We consider a NACA0012 airfoil profile with a cylinder placed underneath and ahead of the leading-edge as our geometry as shown in figure~\ref{fig:domain}. The trailing-edge of the airfoil is slightly rounded ($r_{TE}=0.00128c$) so that there are finite number of Cartesian cells that cover the airfoil trailing edge. The truncated chord length of the airfoil is $c_0=0.924c$, where $c$ is unit chord. For the airfoil-cylinder system, the cylinder of diameter $D=0.12c$ (same as the thickness of the airfoil) is placed at $(x_{c0},y_{c0})$ at vertical gap of $d$ from the airfoil chord. For most of the current simulations, unless specified, $y_{c0}=-1.5D$ and $x_{c0}=x_{LE}-0.5D$. For $D=0.12c$, the cylinder center is placed at $(x_{c0},y_{c0})=(-0.522,-0.18)$. This cylinder placement is considered for a majority of the subsequent analysis unless specified. The computational domain spans 10 chord lengths in all directions with the airfoil as its center. For the earlier mentioned cylinder placement, a uniform mesh with a grid spacing of $\Delta=0.00125c$ is used for $-0.6\leq x/c\leq0.5$ and $-0.6\leq y/c \leq 0.5$ to ensure that that the cylinder-airfoil system and the dominant physics is contained within the uniform, highly resolved region. The grid is stretched outside this uniform region as shown in figure~\ref{fig:domain}(a). The upstream and vertical directions are stretched at the same rate, while the downstream region is stretched more gradually to resolve the wake. Overall, we employ a mesh size of $1240\times 960$ resulting in approximately 1.192 million Cartesian cells. We use the same mesh for the airfoil-only simulations. The grid verification details are discussion in Appendix A.

Both the airfoil only and the airfoil-cylinder systems are pitched about the airfoil mid-chord sinusoidally as per the kinematics given by,
\begin{equation}
    \theta(t) = \theta_0+A_\theta\sin(2\pi f_s t+\pi/2) \label{eq:pitch_motion}
\end{equation}
where $\theta_0$ is set to be $0^\circ$, $A_\theta$ is the pitching oscillation amplitude, frequency is $f_s$ and $T=1/f_s$ is the oscillation period. We set $f_s$ to be fixed value for an appropriate comparison with prior study \citep{Turner2024}. For all the current simulations the reduced frequency is set to $f^\star=f_sc_0/U_\infty=0.1$.

 \section{Transonic flutter characteristics: Prescribed motion simulations}
 \label{sec:flow_physics}
 
 \begin{figure}
\centering
    \includegraphics[width=1.0\textwidth]{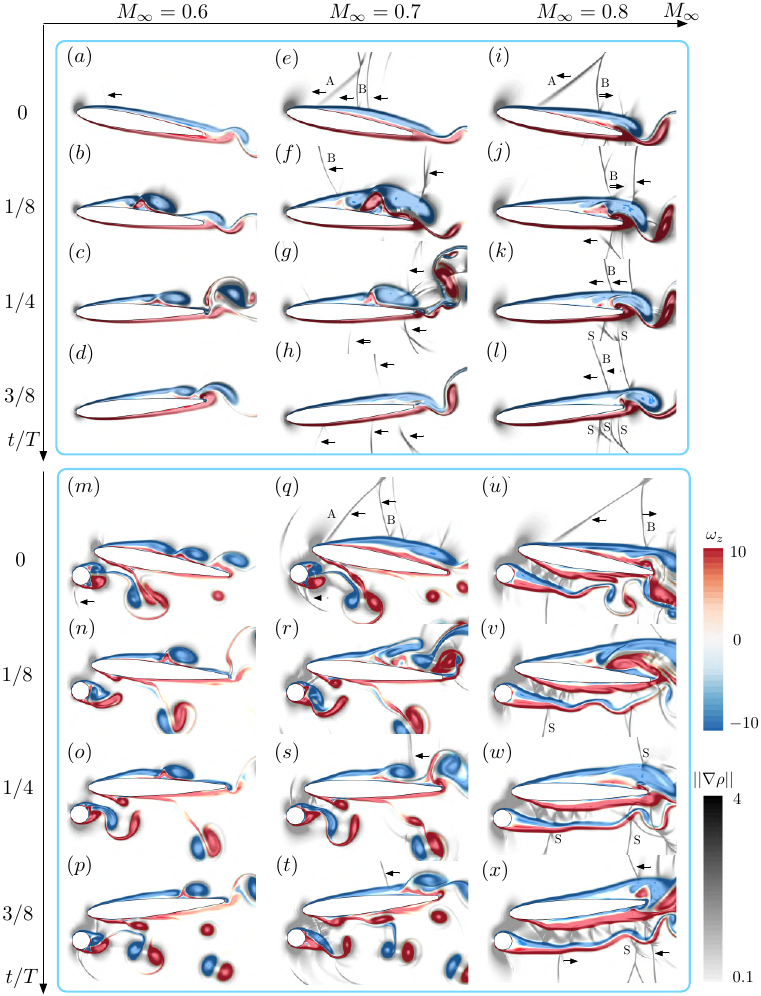}
    \caption{Instantaneous vorticity-fields of a sinusoidally pitching airfoil-only (a)-(l) and cylinder-airfoil systems (m)-(x) with a pitching amplitude of $A_\theta=10^\circ$ and reduced frequency of $f^\star=0.1$. Vorticity-fields are visualized at four instances during the pitching down motion corresponding to $t/T=4,4.125,4.25,4.375$ at Mach numbers $M_\infty=0.6,0.7,0.8$ using $\omega_z$ and $||\nabla \rho||$ in grey. The shock waves are characterized using the arrows indicate the shock motion direction and S indicates a stationary shock.}
    \label{fig:Aoa10_Airfoil_AirfoilCyl}
\end{figure}

 We investigate the transonic flutter characteristics of a two-dimensional sinusoidally pitching NACA0012 airfoil and cylinder-airfoil system under at various transonic Mach numbers and pitching amplitudes at a chord-based Reynolds number of 10,000 with a pitching frequency of $f^\star=0.1$.
 
\subsection{Vortex and Shock Dynamics}

We consider the flow physics of the airfoil-only and the cylinder-airfoil systems pitching with $A_\theta=10^\circ$ at various Mach numbers of $M_\infty=0.6,0.7$ and $0.8$. The corresponding vorticity-fields at four instances during the pitch down motion are visualized in figure~\ref{fig:Aoa10_Airfoil_AirfoilCyl}. Here, $t/T\in[4,4.5]$ captures the pitch-down motion and $t/T\in[4.5,5]$ captures the pitch-up motion. The instance corresponding to the maximum oscillation amplitude is $t/T=4$, where $T$ is the respective time period of oscillation at each inflow Mach number. 

For the pitching airfoil-only system (figure~\ref{fig:Aoa10_Airfoil_AirfoilCyl}(a)-(l)), as Mach number is increased, the flow becomes locally supersonic as can be seen as the shock waves. At $M_\infty=0.6$, while we see weak compression waves at the leading edge at $t/T=4$ (figure~\ref{fig:Aoa10_Airfoil_AirfoilCyl}(a)), the flow remains subsonic throughout the domain. As the airfoil pitches down, we see flow separation and shedding on the upper surface of the airfoil. At $M_\infty=0.7$ (figure~\ref{fig:Aoa10_Airfoil_AirfoilCyl}(e)-(h)), we see multiple upstream propagating shock waves as the airfoil pitches down, indicated by the arrows. These upstream propagating shock waves later dissipate into the freestream and are classified as Tijdeman type C \citep{tijdeman1980transonic}.  At the maximum pitch-up position at $t/T=0$ (figure~\ref{fig:Aoa10_Airfoil_AirfoilCyl}(e)), we see the presence of $\lambda-$shock wave system, where $A$ and $B$ denote the branches of the $\lambda-$shock. Downstream of branch A, we see a separated shear layer. This $\lambda-$ shock causes strong flow separation and shedding downstream of branch $B$ as the airfoil pitches down at $t/T=4.125$ (figure~\ref{fig:Aoa10_Airfoil_AirfoilCyl}(f)). \cite{Turner2024} investigated the flow physics and flutter characteristics for this airfoil-only system with $A_\theta=10^\circ$. They identified the presence of flutter instability at $M_\infty=0.7$ and attributed the flutter mechanism to this  shock-induced separation. The flow separation near the trailing edge causes a low pressure region on the upper surface, creating a negative aerodynamic moment during the pitch-down moment resulting in net positive energy transfer as will be discussed in subsequent sections. At this time instant, a normal shock wave originates at the trailing edge due to the vortex roll-up. The trailing edge vortex undergoes shedding and the normal shock wave propagates upstream on the upper surface as the airfoil pitches down. At $t/T=4.375$  (figure~\ref{fig:Aoa10_Airfoil_AirfoilCyl}(h)), the upstream propagating shock waves on the lower surface later result in the $\lambda-$shock-induced separation during the pitch-up motion. Since the airfoil has zero mean angle of attack, the same physics of the lambda-shock-wave formation and shock-induced stall is observed on the lower surface during the pitch-up motion. 

As the Mach number is further increased to $M_\infty=0.8$ (figure~\ref{fig:Aoa10_Airfoil_AirfoilCyl}(i)-(l)), the width $\lambda-$shock system expands (figure~\ref{fig:Aoa10_Airfoil_AirfoilCyl}(i)) when compared to $M_\infty=0.7$ (figure~\ref{fig:Aoa10_Airfoil_AirfoilCyl}(i)) during the maximum pitch-up position. While the branch A propagates upstream into the freestream, branch B propagates downstream toward the trailing edge of the airfoil and is relatively stationary. As the airfoil starts to pitch down, the shock-induced separation predominantly occurs near the trailing edge and not on the airfoil surface. Past $t/T=4.125$, (figures~\ref{fig:Aoa10_Airfoil_AirfoilCyl}(k) and (l)) both branch B and the normal shock originated due to TE shedding propagate upstream on the upper surface. We also observe the presence of stationary shock waves on the bottom surface of the airfoil.

We now examine the vortex and shock dynamics of the sinusoidally pitching cylinder-airfoil system (figure~\ref{fig:Aoa10_Airfoil_AirfoilCyl}(m)-(x)). With the addition of cylinder, vortex shedding and shear layers now interact with the lower-wing surface aerodynamics. A larger density gradient near the airfoil leading edge now is seen encompassing both the airfoil leading edge, the cylinder and the gap region. At $M_\infty = 0.6$ (figure~\ref{fig:Aoa10_Airfoil_AirfoilCyl}(m)-(p)), unlike the airfoil-only scenario (figure~\ref{fig:Aoa10_Airfoil_AirfoilCyl}(a)-(d)), the flow becomes locally supersonic due to the presence of the cylinder. The presence of cylinder causes flow accelerating in the gap region, where the shear layers on the cylinder and the lower airfoil surface act as a channel, this results in shock waves in the gap region (figure~\ref{fig:Aoa10_Airfoil_AirfoilCyl}(p)). The blockage effect from the cylinder also results in a more pronounced flow separation and shedding on the airfoil upper surface compared to airfoil-only case.

As Mach number is increased to $M_\infty=0.7$, a $\lambda-$shock is formed at $t/T=4$ (figure~\ref{fig:Aoa10_Airfoil_AirfoilCyl}(q)). As compared to the airfoil-only system, we notice the advancement of branch A of the  $\lambda-$shock near the leading edge due to the presence of the cylinder. Similar to the airfoil-only system, flow separation on the airfoil surface is seen downstream of branch B. As seen in $M_\infty=0.6$, even for $M_\infty=0.7$, we observe a pronounced vortex shedding on the upper surface that is also prevalent after the vortex shedding at the trailing edge. Further, the cylinder adds complexity to the flow physics on the airfoil lower surface, resulting in vortex shedding and shock waves in the gap region. As the Mach number is increased to $M_\infty=0.8$, the vortex shedding from the cylinder is suppressed \citep{canuto2015two,liu2023numerical}. The shear layer on the airfoil lower surface and shear layer emanating from the cylinder act as a channel with moving walls. The incoming flow accelerates and becomes supersonic as it enters the gap flow and a shock-train is formed \citep{matsuo2003shock,gnani2016pseudo}. \citet{xiong2018analysis} identified that the shock-train undergoes oscillations due to periodic downstream disturbances in a stationary channel with constant area. In the present case, the gap region continuously is moving and deforming and we also see the flow separation and shedding downstream of the gap region. Hence, the structure of the shock-train continuously changes as the cylinder-airfoil system undergoes the pitching motion. Similar to the airfoil-only case, we observe formation of $\lambda-$ shock waves on the upper airfoil surface in the cylinder-airfoil system. Overall, the addition of cylinder increases the complexity of the flow physics due to the presence of vortex shedding from the cylinder, interaction between the cylinder vortex shedding with the lower airfoil shear layer, flow acceleration due to the blockage effects from the cylinder.  

\subsection{Energy Maps}
The flow physics for both the airfoil-only and cylinder-airfoil systems varies based on the Mach number, pitching amplitude and frequency. In order to systematically identify the flutter boundaries in the Mach number and the pitching amplitude space ($M_\infty-A_\theta$), we consider energy maps. Energy maps have been used to quantify the energy imparted to the structure by the fluid and vice versa for moving bodies in both compressible and incompressible aeroelastic systems \citep{MenonMittal2019,zhu2020nonlinear,turner2024high}. Since we are interested in the effect of underwing cylinder on the airfoil, we consider energy maps solely for the airfoil surface in both the airfoil-only and the cylinder-airfoil systems. For a single degree-of-freedom pitching motion, the energy transfer between the structure and the fluid is quantified as 
\begin{equation}
    E^\star = \int\limits_0^TC_W\,\textrm{d}t,
    \label{eq:Estar}
\end{equation}
where $C_W$ is the power coefficient. The power coefficient is computed as
\begin{equation}
    C_W=C_M\dot{\theta},
    \label{eq:CW}
\end{equation}
where $C_M$ is the pitching moment coefficient and $\dot{\theta}$ is the pitching velocity. In the current scenario, $\dot{\theta}$ is given by the sinusoidal pitching motion prescribed in equation~\ref{eq:pitch_motion}. The pitching moment coefficient is computed from the surface pressure on the airfoil surface $B$ as 
\begin{equation}
    C_M = \frac{\int\limits_B p[(\boldsymbol{x}_p-\boldsymbol{x})\times\boldsymbol{n}]\cdot \boldsymbol{e}_z\,d\boldsymbol{x}}{1/2\rho_\infty U_\infty^2c^2_0},
\end{equation}
where $\boldsymbol{x}_p$ is pivot point (origin in our case), $\boldsymbol{e}_z$ is the unit vector in $z-$direction. The moment arm  is defined such that the pitch-up moment is positive and the pitch-down moment is negative. We note that the contribution of shear stress to $C_M$ is negligible and we compute the moment coefficient based only on the pressure loading on the airfoil. We also note that for the cylinder attached to the wing, the surface pressure from both the surfaces will contribute to the moment but the moment would then depend on the precise shape of the underwing structure and the attachment location. In this study, we are focused on how the aerodynamic interactions of the shear layers, wake, gap flow and the shock dynamics between the airfoil and the underwing cylinder effect the flutter characteristics on the airfoil. We therefore ignore the direct aerodynamic forces on store and their impact on the net loading on wing.

 \begin{figure}
\centering
    \includegraphics[width=1.0\textwidth]{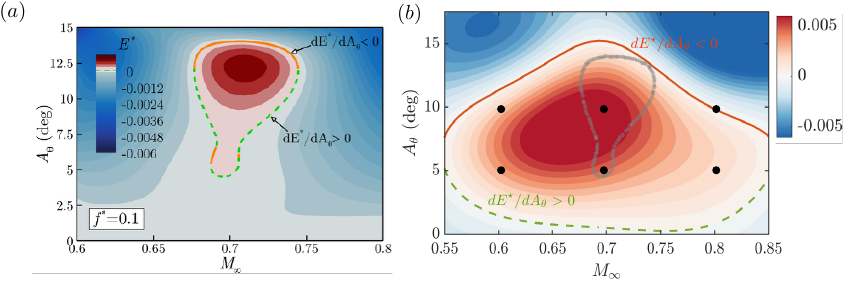}
    \caption{Energy Maps for (a) a pitching airfoil-only system (reproduced from \citep{Turner2024}) and (b) pitching cylinder-airfoil system for Reynolds number of 10,000 and reduced frequency $f^\star=0.1$. The flutter boundary of the airfoil-only system are overlayed in (b) in gray. The black dots denote the flow configurations considered for subsequent analysis.}
    \label{fig:EnergyMap}
\end{figure}

The energy maps for the airfoil-only and the cylinder-airfoil systems in the $M_\infty-A_\theta$ space are shown in figure~\ref{fig:EnergyMap}. The flutter boundaries are characterized by the $E^*=0$ contour. We can classify the flutter boundary as stable or unstable based on the conditions $dE^\star/dA_\theta>0$ and $dE^\star/dA_\theta<0$, respectively \cite{MenonMittal2019}. The energy map for the airfoil-only system is reproduced from \citet{Turner2024} (figure~\ref{fig:EnergyMap}(a)). The energy map for the cylinder-airfoil system is obtained using 70 new simulations sampled in the $M_\infty-A_\theta$ space. For each of these 70 simulations, the energy is computed by averaging over five pitching cycles based on equation~\ref{eq:Estar} after the passage of the initial transients. The energy map is obtained by interpolation using Gaussian process regression (GPR) using the Matern kernel \citep{fan2019robotic}.
\begin{figure}
\centering
    \includegraphics[width=1.0\textwidth]{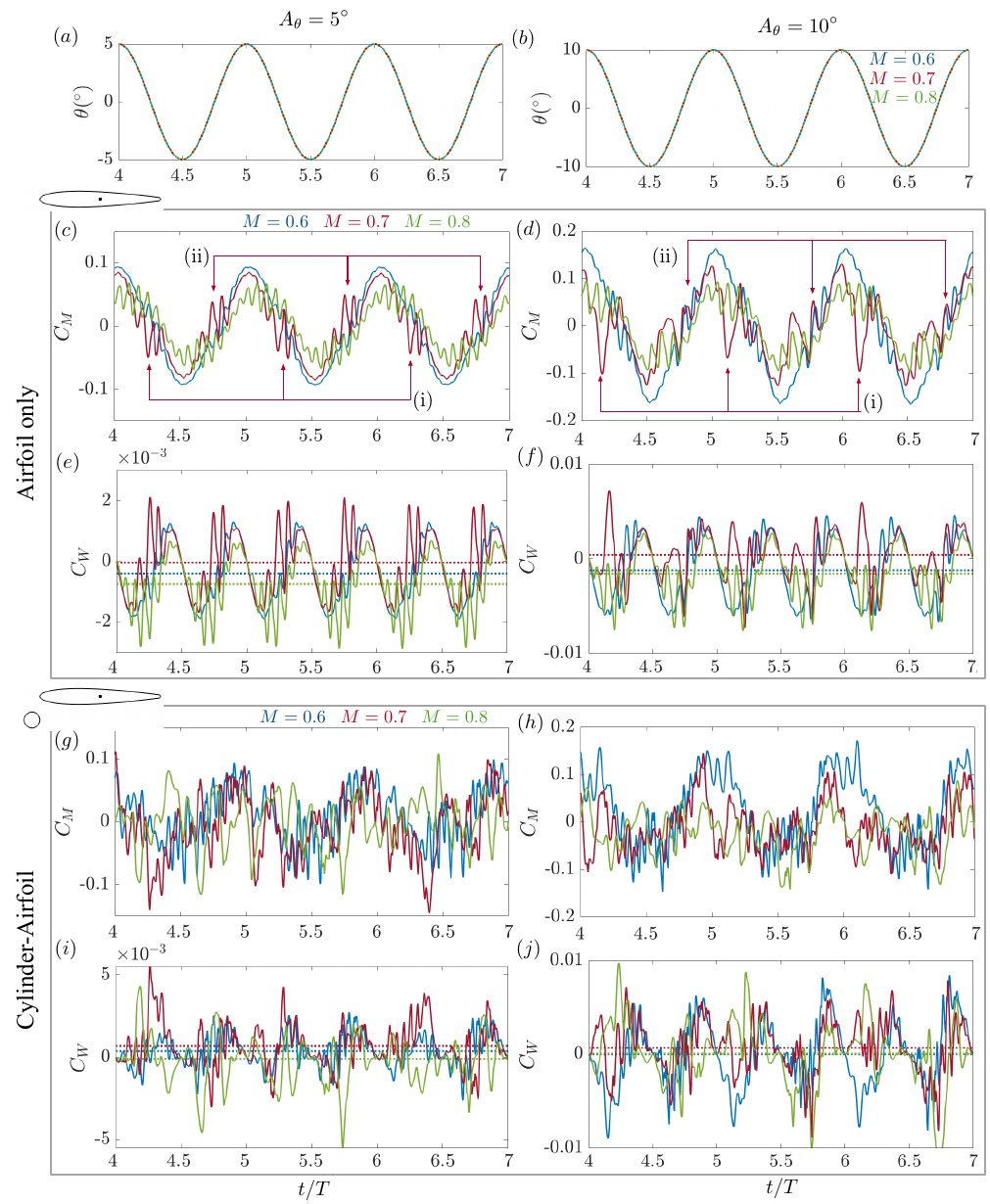}
    \caption{Time history of (a)-(b) the pitching motion $\theta(t)$ and the moment coefficient $C_M$ and the power coefficient $C_W$ of NACA0012 airfoil (c)-(f) and NACA0012 airfoil-cylinder system (g)-(j), respectively at pitching amplitudes of $A_\theta=5^\circ$ and $10^\circ$ and $M_\infty=0.6,0.7$ and 0.8 for three pitching cycles. The dotted lines in (e),(f), (i) and (j) denote the mean power coefficients for each case.}
    \label{fig:CWtime_Airfoil_AirfoilCyl}
\end{figure}

The flutter boundary for the airfoil-only system is overlayed on the energy map of the cylinder-airfoil system using the gray line (figure~\ref{fig:EnergyMap}(b)) and we note that compared to the airfoil-only system, the cylinder-airfoil system is much more prone to flutter instability as the flutter boundaries expand significantly. The maximum energy transfer for the cylinder-airfoil system occurs at $(M_\infty,A_\theta)\approx(0.675,8^\circ)$. Further, the subcritical instability now occurs at significantly lower Mach number of $M_\infty\approx0.56$ for the cylinder-airfoil system compared to the $M_\infty\approx 0.68$ for the airfoil-only system. These initial observations are inline with several previous studies that indicate that the addition of store can enhance transonic flutter \citep{Beran2004,Parker2007,Denegri2013,Iovnovich2017}.

To further examine the effect of the cylinder wake on flutter, we consider several distinct cases corresponding to pitching amplitudes of $A_\theta=5^\circ$ and $10^\circ$ and Mach numbers $M_\infty=0.6,0.7$ and 0.8. From the energy maps (highlighted as black dots in figure~\ref{fig:EnergyMap}(b)), among these flow conditions, 
the airfoil-only system exhibits flutter instability at $(M_\infty,A_\theta)=(0.7,10^\circ)$ and $(M_\infty,A_\theta)=(0.7,5^\circ)$ is right at the flutter boundary. The pitching cylinder-airfoil system is prone to flutter at all these conditions. Hence, these conditions will give us insights into the effect of cylinder flow on the responsible energy transfer mechanisms for flutter instability.

The sinusoidal pitching motion and the temporal variation of $C_M,C_W$ for these cases are shown in figure~\ref{fig:CWtime_Airfoil_AirfoilCyl} for three pitching cycles. For all these cases, the reduced frequency is $f^\star=0.1$. At $(M_\infty,A_\theta)=(0.7,10^\circ)$, the airfoil-only system has significant positive peaks in $C_W(t)$ variation (figure~\ref{fig:CWtime_Airfoil_AirfoilCyl}(f)), responsible for positive energy transfer to the wing. The corresponding negative and positive peaks in $C_M$ during the pitch-down (i) and pitch-up (ii) motion are highlighted using the arrows (figure~\ref{fig:CWtime_Airfoil_AirfoilCyl}(d)). These peaks are associated with the $\lambda$-shock induced vortex shedding at the trailing edge on the upper surface during the pitch-down motion (figure~\ref{fig:Aoa10_Airfoil_AirfoilCyl}(e)-(g)), and on the lower surface on the pitch-up motion respectively. As the pitch amplitude is reduced to $A_\theta=5^\circ$, the fluctuation magnitude of $C_W$ also reduces (figure~\ref{fig:CWtime_Airfoil_AirfoilCyl}(e)). Similar positive peaks of reduced magnitude appear in $C_W$. The associated peaks in $C_M$ are highlighted in figure~\ref{fig:CWtime_Airfoil_AirfoilCyl}(c). As the pitching amplitude reduces at $M_\infty=0.7$, the shock-induced flow separation is suppressed placing $A_\theta=5^\circ$ at the flutter boundary. These peaks are absent in $M_\infty=0.6,0.8$ as they fall outside of flutter boundaries. 

For the pitching cylinder-airfoil system (figures~\ref{fig:CWtime_Airfoil_AirfoilCyl}(g)-(j)), we observe several high frequency fluctuations in $C_M(t)$ unlike the airfoil-only case due to the presence of cylinder wake. To obtain further insights into the variation of $C_W$ and responsible physical mechanisms, we leverage compressible force and power partitioning method.

\section{Compressible Force and Power Partitioning Method}
\label{sec:cfpm}
The force partitioning method (FPM) has been widely applied to understand the flow structures and mechanisms responsible for generating aerodynamic forces and moments in several unsteady aeroelastic and aerodynamic systems \citep{zhang2015centripetal,menon2021initiation,zhu2023flow,prakhar2025modal,kumar2025computational}. This method is based on the work by \citet{quartapelle1983force} and and is also related to the formulation of \citet{Chang1992}. FPM partitions the aerodynamic forces and moments into volume and surface integrals on an immersed body by projecting the momentum conservation equation on an influence influence field. The details of the FPM method for incompressible flows can be found in \cite{zhang2015centripetal,menon2021initiation}.

In this study, we extend this force partitioning framework to \emph{compressible} flows. The related influence-potential based force element method using the lamb-vector has been proposed in the past  \citet{ChangLei1996,ChangSuLei1998} for compressible flows past stationary bodies to examine the volumetric components for the aerodynamic forces. However, extension to moving bodies in compressible flows has not been shown before. Furthermore, since energy-transfer between the flow and a oscillating airfoil is not just driven by the pressure loading but also the phase between the pressure loading and the airfoil motion, application of partitioning directly to the power associated with flutter is needed. Using these extensions, we can investigate the flow physics responsible for transonic wing flutter and the effect of wake interaction in the presence of a cylinder. Since energy is a quantity of interest for flutter, in the present study, we solely focus on the partitioning of power. 

\begin{figure}
\centering
    \includegraphics[width=0.9\textwidth]{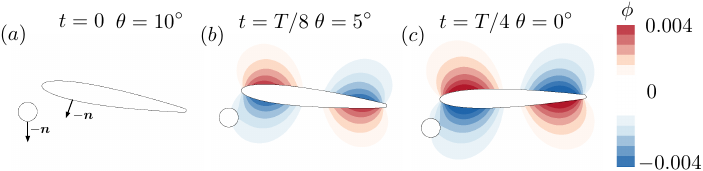}
    \caption{ Influence fields corresponding to power $(\phi)$ (a)-(c) for a pitching cylinder-airfoil system with $f^\star=0.1$ and $A_\theta=10^\circ$ at three different time instances of a pitching cycle. Note that $\phi=0$ at $t=0$ as the body velocity is zero at the maximum pitch-up phase.}
    \label{fig:phi}
\end{figure}

For partitioning of induced power, we start by constructing the ``influence'' field $\phi(\boldsymbol{x})$ defined as follows: 
\begin{equation}
  \nabla^2 \phi = 0,
\qquad
\text{with}
\qquad
\boldsymbol{n}\cdot\nabla \phi =
\begin{cases}
\boldsymbol{n\cdot v_B}, & \text{on } B, \\
0,   & \text{on } \Sigma,
\end{cases}  
\label{eq:aux_potential}
\end{equation}
where $\boldsymbol{n}$ is the unit normal vector on the immersed body surface ($B$) (the normals for the airfoil and the cylinder can be seen in figure~\ref{fig:phi}(a)) and the outer domain boundaries $(\Sigma)$, and $\boldsymbol{v_B}$ is the velocity of the immersed surface. Note that this is subtly different from the influence potential corresponding to the aerodynamic forces or moments \citep{menon2021initiation} in that the velocity of the body surface $\boldsymbol{v_B}$ also appears in the construction of the influence potential. A few instances of the influence fields corresponding to power for a sinusoidally pitching cylinder-airfoil system are visualized in figures~\ref{fig:phi}(a)-(c) and (d)-(f), respectively. Note that the presence of the cylinder changes the influence field in a small but noticeable manner. 

By projecting the compressible momentum equation (equation ~\ref{eq:NS_comp}) on $\nabla \phi$ and integrating over the fluid volume $\mathcal{V}$ we obtain,
\begin{align}
    \int_{\mathcal{V}}\nabla p \cdot \nabla\phi dV = \int_{\mathcal{V}}\left[\nabla\cdot\boldsymbol{\tau}-\frac{\partial \rho \boldsymbol{u}}{\partial t} - \nabla \cdot (\rho \boldsymbol{u}\otimes \boldsymbol{u})\right]\cdot\nabla\phi\, dV.
    \label{eq:cFPM_a}
\end{align}
Using Gauss-divergence theorem, conservation of mass and the boundary conditions for influence potential (equation~\ref{eq:aux_potential}), we can further simplify the equation~\ref{eq:cFPM_a} into equation~\ref{eq:cFPM_b}. 
\begin{align}
    \int_{B}p\boldsymbol{n}\cdot\boldsymbol{v_B}dS = \int_{\mathcal{V}}\left[\nabla \cdot \left(\rho\frac{D\boldsymbol{u}}{Dt}\right)\phi + (\nabla\cdot \boldsymbol{\tau})\cdot\nabla\phi\right]dV+\int_{B+\Sigma}\phi\left[-\rho\boldsymbol{n}\cdot{\frac{D\boldsymbol{u}}{Dt}}\right]dS.
    \label{eq:cFPM_b}
\end{align}
Since the outer boundaries $(\Sigma)$ are placed far away from the immersed body where the flow is mostly uniform in space and time, we can simplify the surface integral to finally obtain the partitioned power as
\begin{equation}
\begin{aligned}
    \underbrace{\int_{B}p\boldsymbol{n}\cdot\boldsymbol{v_B}dS}_{\mathbf{W}} = \underbrace{\int_{\mathcal{V}}\phi\left[\nabla \cdot \left(\rho\frac{D\boldsymbol{u}}{Dt}\right)\right]\mathrm{d}V}_{\mathbf{W}_1}+\underbrace{\int_{\mathcal{V}}(\nabla\cdot\boldsymbol{\tau})\cdot\nabla\phi\,\mathrm{d}V}_{\mathbf{W}_2}+
     \underbrace{\int_{B}\phi\left[-\rho\boldsymbol{n}\cdot{\frac{D\boldsymbol{u}}{Dt}}\right]\mathrm{d}S}_{\mathbf{W}_3}.
\end{aligned}
\label{eq:cPPM_split}
\end{equation}
The time-integrated power per one time-period of oscillation amounts to the energy exchanged between the flow and the foil. The respective integrands are denoted as  $\mathbf{W}_j=\int\mathbf{w}_j(\boldsymbol{x}) \mathrm{d}\boldsymbol{x}$ where the subscript $j$ corresponds to the specific partition. From equation \ref{eq:cPPM_split}, we have two volume integrals and one surface integral. The integrand of the first term on the right hand side $(\mathbf{W}_1)$ corresponds to the volumetric Lagrangian rate of change of momentum that encompasses the volumetric unsteady and advection effects. The term $\mathbf{W}_2$ represents the volumetric effects of viscous stress divergence on surface pressure. The third right hand side integrand $(\mathbf{W}_3)$ encodes the added-mass effects which depends only on the surface acceleration of the immersed body. 

The power partitioning framework for compressible flows can be directly compared with the corresponding formulation for incompressible flows \citep{zhang2015centripetal,menon2021initiation}. The terms $\mathbf{W}_2$ and $\mathbf{W}_3$ appear in both formulations. In the incompressible limit, the viscous stress divergence term, $\mathbf{W}_2$, reduces to the viscous momentum diffusion term because the velocity divergence vanishes \citep{aghaei2022contributions}. The primary distinction between the incompressible and compressible formulations arises in the $\mathbf{W}_1$ term. For incompressible flows, this term reduces to the so-called ``vortex-induced force'', whose integrand contains $Q=-\nabla \cdot \left({\boldsymbol{u}\otimes \boldsymbol{u}} \right)$. In contrast, the compressible formulation retains the divergence of the full Lagrangian rate of momentum change, thereby incorporating not only the nonlinear convective contribution, but also dilatational and density-gradient effects that are absent in incompressible flows. Consequently, $\mathbf{w}_1$ represents the combined influence of vortical motion, compressibility-induced dilatation, and density gradients on the surface pressure loading.

 \begin{figure}
\centering
    \includegraphics[width=1.0\textwidth]{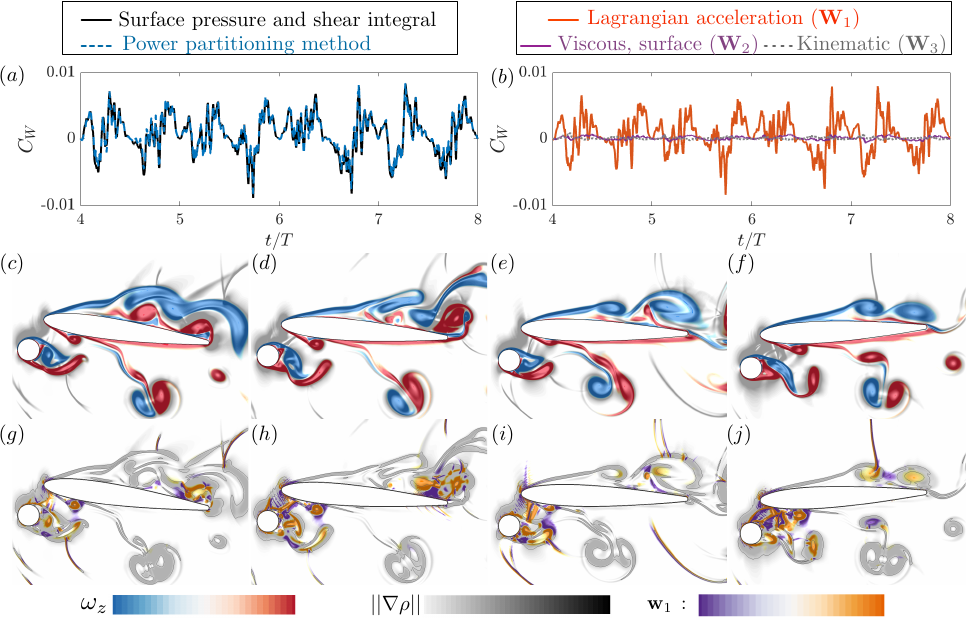}
    \caption{Application of compressible power partitioning method to the pitching cylinder-airfoil system at $M_\infty=0.7$ and $A_\theta=10^\circ$. (a) Verification of the power partitioning method. (b) Contributions to $C_W$ from individual power partitions. (c)-(f) A few instances of flowfields visualized using vorticity contours and density gradient magnitude in gray. (g)-(j) The corresponding power density fields colored using $\mathbf{w}_1$, $||\nabla \rho||$ in gray and accentuated using $||\nabla \rho||=3$ contour line.}
    \label{fig:cppm_verify}
\end{figure}

The application of the power partitioning method to the pitching cylinder-airfoil system at $(M_\infty,A_\theta=0.7,10^\circ)$ is shown in figure~\ref{fig:cppm_verify} as an example. Figure~\ref{fig:cppm_verify}(a) shows that $C_W$ computed using the power partitioning method (equation~\ref{eq:cPPM_split}) matches with the convention computation using the surface integral of power. The contributions from the individual terms can be seen in figure~\ref{fig:cppm_verify}(b). The Lagrangian rate of change of momentum term, $\mathbf{W}_1$, is the primary energy transfer term responsible for about 85\% of total energy, followed by the volumetric viscous term $\mathbf{W}_2$, which accounts for 14\%. The added mass term, $\mathbf{W}_3$, contribution to the overall energy transfer is negligible. The spatial regions of the flow which have non-zero contributions to the energy transfer can be directly visualized using $\mathbf{w}_1(\boldsymbol{x},t)$ figure~\ref{fig:cppm_verify}(g)-(j). The cylinder wake, the gap flow, shear layers on airfoil surface and the vortex shedding near the trailing edge are highlighted. We also note that the shock waves are decomposed as both positive and negative energy transfer due to the pressure gradient across the shock waves. Furthermore, the contribution of structures diminishes with distance from the airfoil due to the rapid decay of the influence field.

\subsection{Quadrant-wise Regional Power Partitioning}
\begin{figure}
\centering
    \includegraphics[width=1.0\textwidth]{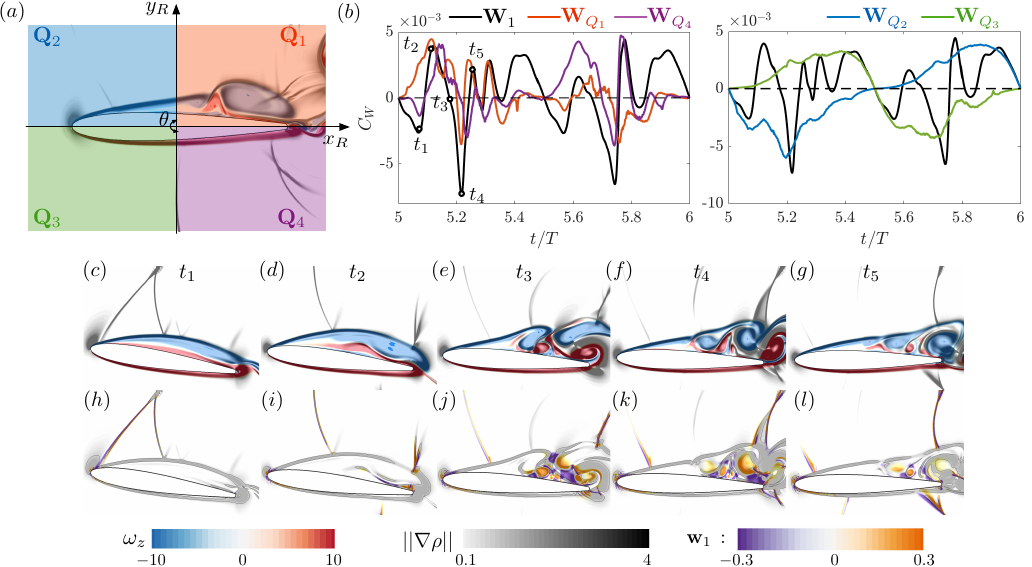}
    \caption{Application of regional power partitioning method for flow over a pitching NACA0012 airfoil at $M_\infty=0.7$ and $A_\theta=10^\circ$. (a) Illustration of quadrant-based splitting of power. (b) Contributions to instantaneous $C_W$ from individual quadrants for one pitching cycle. (c)-(g) A few instances of flowfields visualized using $\omega_z$ and density gradient magnitude in gray. (h)-(l) The corresponding power density fields colored using $\mathbf{w}_1$ and $||\nabla \rho||$ in the background in gray along with $||\nabla \rho||=3$ contour line.}
    \label{fig:Rppm_airfoil_01}
\end{figure}

Splitting the integral into different regions of interest can help isolate the mechanism of flutter instability \citep{menon2021initiation}. Here, we split the integrand into four distinct quadrants: $\boldsymbol{Q}_k$, $k =$ 1, 2, 3, 4. These are centered around the mid-chord pivot point of the airfoil (see figure~\ref{fig:Rppm_airfoil_01}(a)) and rotate with the airfoil as the airfoil pitches up and down. Specifically, we consider body-fitted coordinates denoted as $\boldsymbol{x}_R(t)$ centered around the origin. For a sinusoidally pitching system about the origin, these can be obtained as
\begin{equation}
\begin{bmatrix} 
x_R(t)\\
y_R(t)
\end{bmatrix}=\begin{bmatrix}
    \cos(\theta(t)) &-\sin(\theta(t))\\
    \sin(\theta(t)) &\cos(\theta(t)).
\end{bmatrix}\begin{bmatrix}
    x\\y
\end{bmatrix}
\label{eq:Quad_rot}
\end{equation}

% \begin{figure}
%\centering
%    \includegraphics[width=0.9\textwidth]{figs/mfig_AirfoilM7A10_RPPM02_v2.pdf}
%    \caption{(a) Phase-averaged and time-reflected power coefficient contributions from the leading edge ($\tilde{\mathbf{W}}_{Q_2}+\tilde{\mathbf{W}}_{Q_3}$) and the trailing edge ($\tilde{\mathbf{W}}_{Q_1}+\tilde{\mathbf{W}}_{Q_4}$) quadrants and (b) Histogram of quadrant-wise contribution to energy for a pitching NACA0012 airfoil corresponding to $M_\infty=0.7$ and $A_\theta=10^\circ$.}
%    \label{fig:Rppm_airfoil_02}
%\end{figure}
 We can then determine the contribution to the power coefficient from each quadrant as,
\begin{equation}
    \mathbf{W}_{\mathbf{Q}_k}(t) = \int_{\boldsymbol{x}_R\in\mathbf{Q}_k}\mathbf{w}_1(\boldsymbol{x},t)\mathrm{d}\boldsymbol{x}. 
\end{equation}
This quadrant-wise division is natural since it allows us to separate the contributions from the upper and lower surfaces of the airfoil and also the leading and trailing edge regions. 
We now apply the regional power partitioning method the case corresponding to $(M_\infty,A_\theta)=(0.7,10^
\circ)$ as it is within the flutter boundary (gray line in figure~\ref{fig:EnergyMap}(b)) and has been investigated in the prior study. The instantaneous quadrant-wise contribution to $C_W$ is seen in figure~\ref{fig:Rppm_airfoil_01}(b). The contributions from quadrants 2 and 3 (figure~\ref{fig:Rppm_airfoil_01}(b) right) vary from positive to negative during the pitch-up and pitch-down alternatively. The power density fields in $Q_2$ and $Q_3$ highlights the leading edge shear region and the shock waves (figure~\ref{fig:Rppm_airfoil_01}(h)-(l)).  During the pitch-down motion $(t/T\in[0,0.5])$, the upper surface shear layer ($Q_2$) contributes negative energy to the foil. The leading-edge shear layer results in a lower pressure on the suction surface creating a positive aerodynamic moment. Since this occurs during the pitch-down motion $(\dot{\theta}<0)$, we see a negative energy transfer in $Q_2$ and similarly, a positive energy transfer during the pitch-up motion. 

During the pitch down motion, $\mathbf{W}_{Q_1}$ follows the fluctuations in the total $\mathbf{W}_1$ including the significant positive peak at $t_2$. A few instances of the flow fields and the corresponding power density fields during the pitch-down motion can be seen in figure~\ref{fig:Rppm_airfoil_01}(c)-(l).  The flow separation induced by $\lambda$-shock system near the upper surface trailing edge results in the positive peak in $W_{Q_1}$ at $t_2$. Through $\mathbf{w}_1(\boldsymbol{x},t)$, we can also identify which region of flow separation results in positive or negative energy transfer. The trailing-edge vortex shedding process comprises of both positive and negative fluctuations in $W_{Q_1}$ due to the presence of multiple co- and counter-rotating vortices. For instance, the vortex cores at $t_4$ are highlighted with positive $\mathbf{w}_1$ but the region between then is highlighted with negative $\mathbf{w}_1$. Similarly, during the pitch-up motion $t/T\in[0.5,1]$, $\mathbf{W}_{Q_4}$ explains the positive peak in $\mathbf{W}_1$ at $t/T\approx0.8$. Hence, regional power partitioning reconfirms that the shock-induced flow separation is responsible for flutter in the airfoil-only case. 

%For the airfoil-only case, there is an inherent symmetry between the pitch-up and pitch-down motion in terms of the contributions of the flow to the energy transfer mechanisms. However, this symmetry is disrupted due to the presence of the cylinder in the cylinder-airfoil case. We examine this symmetry-breaking phenomena by considering the  time-reflected symmetry in $\theta(t)$ given by
%\begin{equation}
%     \theta(t)=-\theta(T/2+t), \quad t\in[0,T].
%     \label{eq:time_reflect}
% \end{equation}
% We can then define net time-reflected aerodynamic coefficients to highlight the asymmetric component of the pitching dynamics as
% \begin{equation}
%     \tilde{\chi}(t)=\chi(t)+\chi(T/2+t), \quad t\in[0,0.5T],
%     \label{eq:time_reflect}
% \end{equation}
% where $\chi$ can be any aerodynamic coefficient and the quadrant-wise variations. The time-reflected coefficient is the sum at two instants in a cycle corresponding to the equivalent phases during the pitch-down and pitch-up motions.

 \section{Influence of cylinder wake on airfoil flutter}
 \label{sec:cylinder_influence}
\begin{figure}
\centering
    \includegraphics[width=1.0\textwidth]{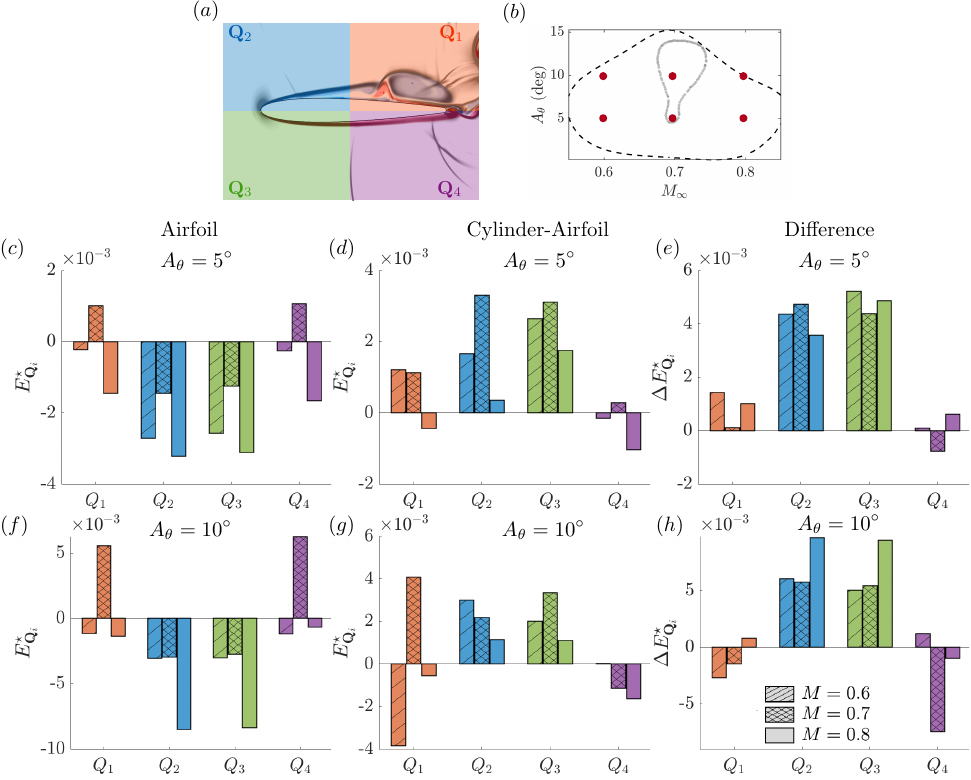}
    \caption{Quadrant-wise contribution of energy transfer $E^\star_{Q_i}$. (a) Schematic of color-coded quadrant division and (b) Outline of flutter boundaries for airfoil-only (gray) and cylinder-airfoil (black dashed) systems. The cases considered in (c)-(h) are shown as red dots.  $E^\star_{Q_i}$ for airfoil-only ((c),(f)), cylinder-airfoil systems ((d),(g)) and the difference between them ((e),(h)) for $A_\theta=5^\circ$ and $10^\circ$ and $M_\infty=0.6,0.7$ and 0.8.}
    \label{fig:Rppm_hist}
\end{figure}
 We apply the regional power partitioning method for airfoil-only and cylinder-airfoil systems to compare the flow physics and the energy transfer mechanisms at the pitching amplitudes of $A_\theta=5^\circ$ and $10^\circ$ and Mach numbers $M_\infty=0.6,0.7$ and 0.8. From the energy maps (figure~\ref{fig:EnergyMap}), among these flow conditions, 
the airfoil-only system exhibits flutter instability at $(M_\infty,A_\theta)=(0.7,10^\circ)$ and $(M_\infty,A_\theta)=(0.7,5^\circ)$ is right at the flutter boundary. Comparatively, the cylinder-airfoil system is more prone to flutter at all these conditions. The overall energy contribution can be quantified using the quadrant-wise energy contribution $E^\star_{Q_i}$ in figure~\ref{fig:Rppm_hist}. Here, $E^\star_{Q_i}=\int\limits_0^T W_{Q_i}dt$ and $i\in\{1,2,3,4\}$. The difference in the energy transfer due to the presence of the cylinder is computed using $\Delta E^\star_{Q_i}$ (figure~\ref{fig:Rppm_hist}(c), (f)) given as,
\begin{equation}
    \Delta E^\star_{Q_i} = {E^\star_{Q_i}}^{\textrm{cylinder-airfoil}}-{E^\star_{Q_i}}^{\textrm{airfoil-only}},\; i\in\{1,2,3,4\}.
    \label{eq:deltaE}
\end{equation}

\begin{figure}
\centering\includegraphics[width=0.95\textwidth]{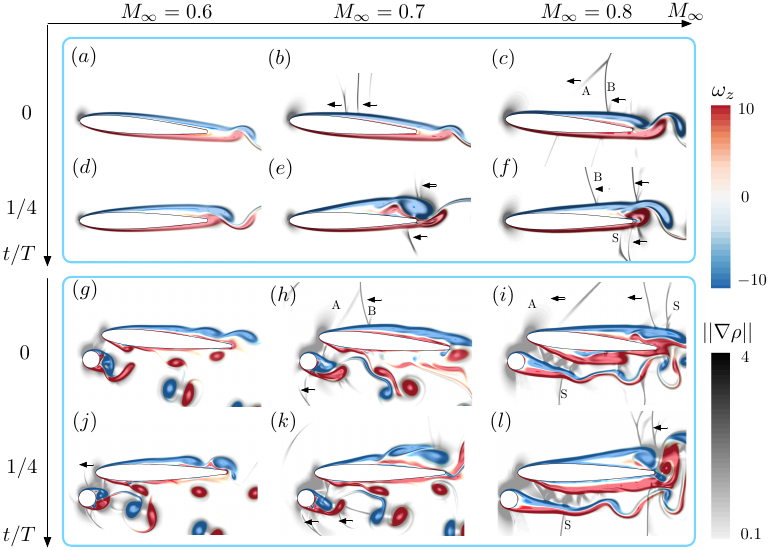}
    \caption{Instantaneous vorticity-fields of a sinusoidally pitching airfoil-only (a)-(f) and cylinder-airfoil (g)-(l) systems with a pitching amplitude of $A_\theta=5^\circ$ are visualized using $\omega_z$ and $||\nabla \rho||$ corresponding to $t/T=4,4.25$ at $M_\infty=0.6, 0.7$ and 0.8. The arrows indicate the instantaneous shock motion direction and S indicates a stationary shock.}
    \label{fig:AoA5_flow}
\end{figure}
\begin{figure}
\centering
    \includegraphics[width=1.0\textwidth]{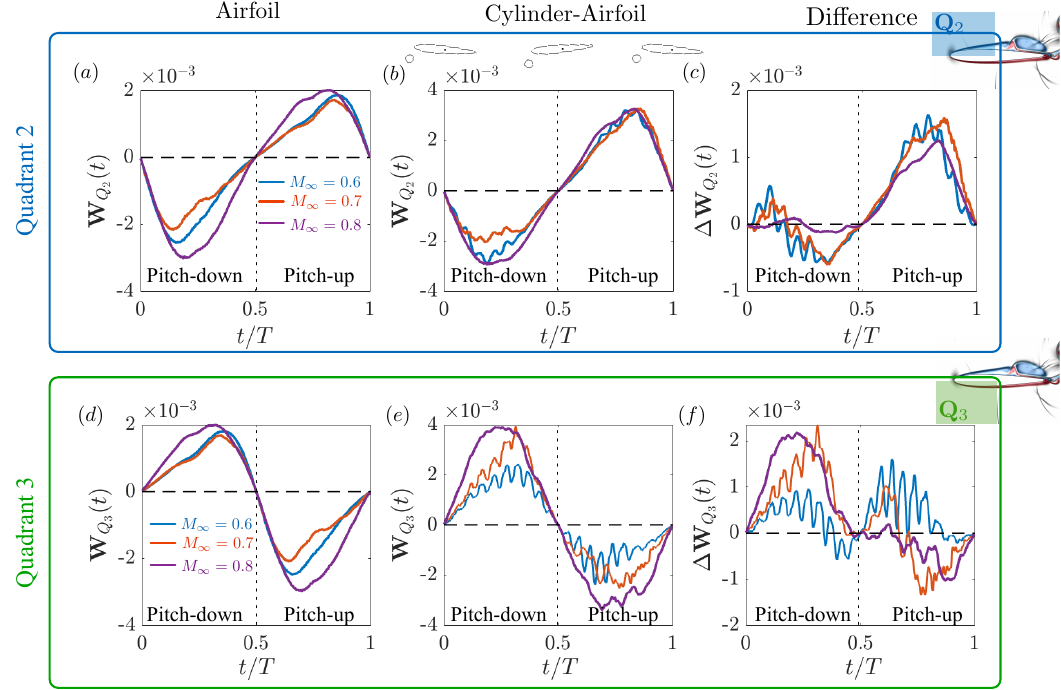}
    \caption{Dominant effects of underwing cylinder on airfoil flutter. Power coefficient contribution from (a)-(c) $Q_2$, and (d)-(f) $Q_3$  at $A_\theta=5^\circ$ and $M_\infty=0.6,0.7$ and 0.8. Phase-averaged  $\mathbf{W}_{Q_2}$ and $\mathbf{W}_{Q_3}$ for (a),(d) airfoil-only system, (b),(e) cylinder-airfoil system and (c),(f)  $\Delta \mathbf{W}_{Q_2}$ and $\Delta \mathbf{W}_{Q_3}$ are the respective differences between them.}
    \label{fig:CW_Q21}
\end{figure}
For the pitching airfoil-only case (figure~\ref{fig:Rppm_hist}(c),(f)), $M=0.7$ shows flutter instability where the positive energy transfer is seen in $Q_1$ and $Q_4$ for both $A_\theta=5^\circ$ and $10^\circ$. The shock-induced stall and trailing edge shedding is only present at $M=0.7$. The pitching amplitude, $A_\theta=5^\circ$ lies on the flutter boundary and  the net energy reduces compared to $A_\theta=10^\circ$. When the underwing cylinder is introduced in $Q_3$ (figure~\ref{fig:Rppm_hist}(d),(g)), $Q_2$ and $Q_3$ show net energy extraction from the fluid for both pitching amplitudes and at all the Mach numbers. With the cylinder present , similar to the airfoil-only case, $Q_1$ contributes to positive energy transfer for both $A_\theta=5^\circ$ and $10^\circ$ at $M_\infty=0.7$ (figure~\ref{fig:Rppm_hist}(d),(g)). This indicates a similar mechanism of $\lambda-$ shock-induced flow separation on the upper airfoil surface. The corresponding vorticity-fields can also be seen in figure~\ref{fig:Aoa10_Airfoil_AirfoilCyl}(q)-(t). However, at $A_\theta=10^\circ$ (figure~\ref{fig:Rppm_hist}(g)), $Q_4$ contributes to negative energy for $M=0.7$. The presence of cylinder creates vortex shedding on the lower airfoil surface and limits the formation of $\lambda-$shock system (see figure~\ref{fig:Aoa10_Airfoil_AirfoilCyl}(q)-(t)). 

The influence of cylinder flow on the energy transfer can be isolated by considering $\Delta E^\star_{Q_i}$ (equation~\ref{eq:deltaE}) (figure~\ref{fig:Rppm_hist}(e), (h)) where the dominant effects are in the leading-edge quadrants, $Q_2$ and $Q_3$. At $A_\theta=5^\circ$ the addition of upstream cylinder results in an increased energy transfer in all quadrants for all Mach numbers except for $M_\infty=0.7$ in $Q_4$. This discrepancy can be explained by the suppression of $\lambda-$shock system on the airfoil lower surface. 

To further investigate the responsible mechanisms and the dominant effects due to the addition of cylinder, we consider $A_\theta=5^\circ$ and $M_\infty=0.6,0.7$ and 0.8 for the subsequent analysis.
The instantaneous vorticity-fields during the pitch-down motion at $A_\theta=5^\circ$ are visualized in figure~\ref{fig:AoA5_flow} and we note that the flow behavior follow similar trends as $A_\theta=10^\circ$ in figure~\ref{fig:Aoa10_Airfoil_AirfoilCyl}, except for $M_\infty=0.6$. While the airfoil-only system exhibits flow separation on the airfoil surface at $A_\theta=10^\circ$, at $A_\theta=5^\circ$, flow separation occurs at the trailing edge. In contrast, the addition of the cylinder causes flow separation on the airfoil surface and the gap region results in flow acceleration to supersonic speeds. 

Since the quadrants $Q_2$ and $Q_3$ have a dominant effect on enhancing flutter in the cylinder-airfoil case, (see figure~\ref{fig:Rppm_hist}(b) and (e)) we now focus on a detailed analysis of these quadrants.  Quadrants $Q_1$ and $Q_4$, which provide a much smaller contribution to the flutter enhancement are included in Appendix \ref{appB}.

%\subsection{Dominant effects of underwing cylinder on airfoil flutter}
\subsection{Quadrants $Q_2$ and $Q_3$}
The contributions from $Q_2$ and $Q_3$ to $C_W$ for the airfoil and the cylinder-airfoil systems ($\mathbf{W}_{Q_2}$, $\mathbf{W}_{Q_3}$), and their difference ($\Delta \mathbf{W}_{Q_2}$, $\Delta \mathbf{W}_{Q_3}$) are shown in figure~\ref{fig:CW_Q21}. In $Q_2$ (figure~\ref{fig:CW_Q21}(a)-(b)), the energy extraction to the structure from fluid is observed during the pitch-up phase for all three Mach numbers for both the systems. As the Mach number increases from 0.6 to 0.8, we see a suppression of high-frequency fluctuations in $\mathbf{W}_{Q_2}$ due to the suppression of cylinder vortex shedding as can be seen in figure~\ref{fig:AoA5_flow}. The difference in $\mathbf{W}_{Q_2}$ from the airfoil-only system is visualized as $\Delta \mathbf{W}_{Q_2}$ in figure~\ref{fig:CW_Q21}(c). The variation of $\Delta \mathbf{W}_{Q_2}$ for all three Mach numbers show a similar trend. While both the cylinder–airfoil and airfoil-only systems exhibit similar energy transfer during the pitch-down phase, the cylinder–airfoil configuration produces substantially greater energy transfer to the foil during the pitch-up phase ($t/T \in [0.5,1]$). 

Figure~\ref{fig:CW_Q21}(d)-(e), show the phase-averaged $\mathbf{W}_{Q_3}$ for the airfoil-only and the cylinder-airfoil systems, and a significant influence of cylinder can also be seen in this quadrant, which contains the underwing cylinder. While the pitch-up phase is the energy extraction phase in $Q_2$, in $Q_3$, the pitch-down phase is the energy extraction phase for both the systems. Similar to $Q_2$, for the cylinder-airfoil system, the high frequency fluctuations are reduced with increasing $M_\infty$ due to the suppression of cylinder wake shedding (figure~\ref{fig:CW_Q21}(e)). At $M_\infty=0.6$ and 0.7, the high frequency fluctuations are significant in ${\mathbf{W}}_{Q_3}$ (figure~\ref{fig:CW_Q21}(e)) compared to ${\mathbf{W}}_{Q_2}$ (figure~\ref{fig:CW_Q21}(b)) as the cylinder wake shedding directly affects the flow physics in $Q_3$. The difference from the airfoil-only system $\Delta \mathbf{W}_{Q_3}$ (figure~\ref{fig:CW_Q21}(f)) shows an increased energy transfer to the airfoil due to the cylinder during the pitch-down phase for all $M_\infty$. Interestingly, at $M_\infty=0.6$, the cylinder-airfoil system exhibits a significant reduction in the negative energy transfer to the foil during during the pitch-up phase (seen as a positive $\Delta\mathbf{W}_{Q_3}$ in figure~\ref{fig:CW_Q21}(f)), indicating that the presence of the cylinder effects both the pitch-up and pitch-down phases of the foil. Both $M_\infty=0.7$ and 0.8, show a similar trend in $\Delta\mathbf{W}_{Q_3}$, wherein the presence of the cylinder significantly increases the energy transfer to the foil during the pitch-down phase ($t/T \in [0,0.5]$), with a slightly negative contribution during the pitch-up phase.

\begin{figure}
\centering
    \includegraphics[width=1.0\textwidth]{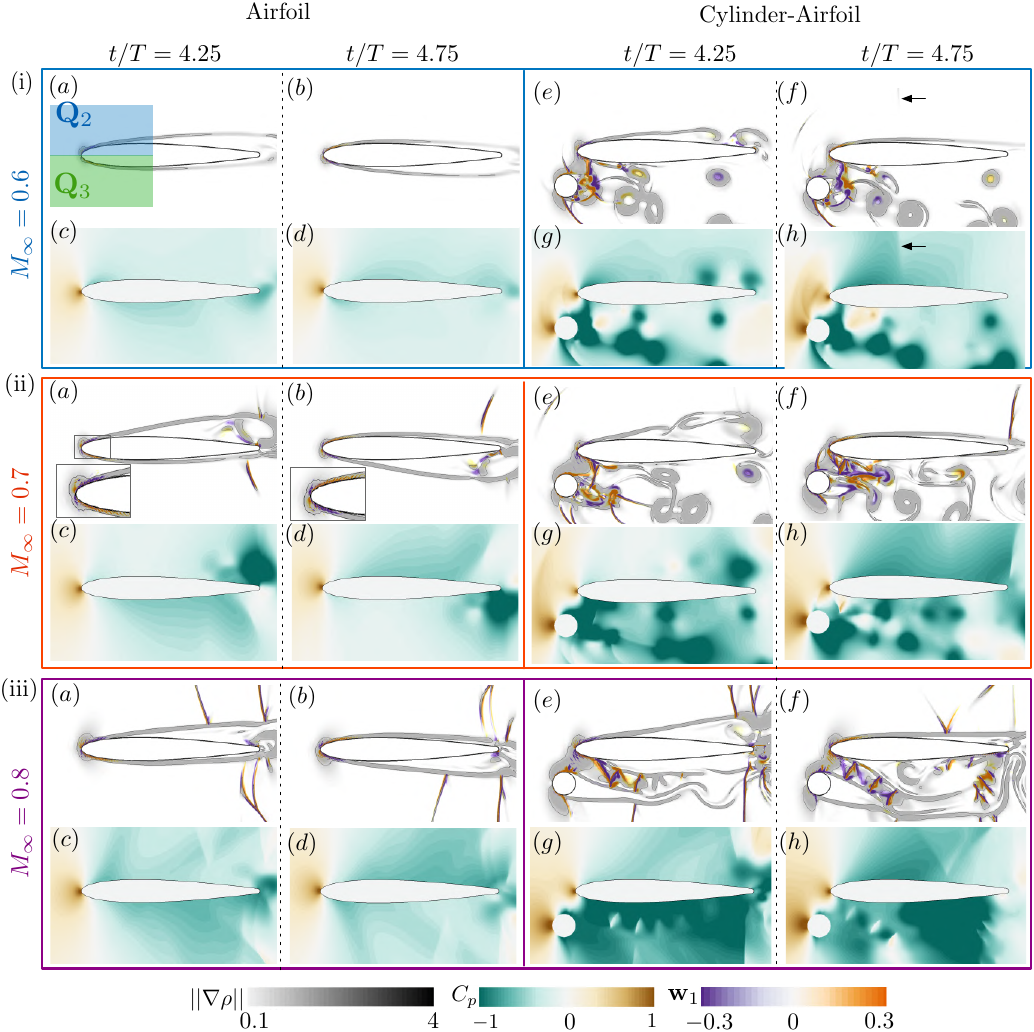}
    \caption{Dominant effect of underwing cylinder on airfoil flutter for $A_\theta=5^\circ$ and $M_\infty=0.6$ (i),0.7 (ii) and 0.8 (iii). Power density fields visualized using $\mathbf{w}_1$, $||\nabla \rho||$ and $||\nabla \rho||=3$ iso-contour (a),(b),(e),(f) and the corresponding pressure coefficient fields $C_p=(p-p_\infty)/(1/2\rho_\infty M_\infty^2)$ (c),(d),(g),(h) at instances $t/T=4.25$ (pitch-down) and 4.75 (pitch-up). Zoomed-in insets near the leading edge are shown in (ii)-(a),(b).}
    \label{fig:CW_Q22}
\end{figure}

\subsubsection{Flow Features and Pressure Distribution}
We now examine the flow-fields to further analyze the influence of cylinder in $Q_2$ and $Q_3$. Figure~\ref{fig:CW_Q22} shows the power density fields, ($\mathbf{w}_1(\boldsymbol{x},t)$ with numerical Schlieren ($||\nabla \rho||$) in the background, and the corresponding pressure coefficient fields $\left(C_p=\frac{p-p_\infty}{1/2\rho_\infty U_\infty^2}\right)$ for $M_\infty=0.6,0.7$ and 0.8 for both systems at $t/T=4.25$ (pitch-down) and 4.75 (pitch-up). The pressure-coefficient fields show the direct impact of the flow structures on the airfoil surface and can be attributed to power, since $C_M$ is computed based on the airfoil surface pressure. 

For the airfoil-only system, in $Q_2$ and $Q_3$, $\mathbf{w}_1(\boldsymbol{x},t)$ is concentrated near the leading-edge shear layers (see the insets in figure~\ref{fig:CW_Q22}(ii)-(a),(b)). Similarly, for the cylinder-airfoil system (figure~\ref{fig:CW_Q22}(i--iii)-(e,f)), $\mathbf{w}_1(\boldsymbol{x},t)$ in $Q_2$ again highlights the leading edge shear layer. In contrast, the quadrant $Q_3$ for the cylinder-airfoil system (figure~\ref{fig:CW_Q22}(i--iii)-(e,f)), highlights not only the shear layers on the airfoil lower surface, but also the shear layers on the cylinder surface, the features in the cylinder wake and the gap region between the airfoil and the cylinder. 

From figure~\ref{fig:CW_Q21}(a)-(b), we identified that the pitch-up phase is the energy extraction phase in $Q_2$ for both airfoil-only and the cylinder-airfoil systems. For both these systems, during the pitch-up phase ($t/T=4.75$) (figure~\ref{fig:CW_Q22}(i--iii)-(d),(h)), $C_p$ field shows a low pressure region on the airfoil upper surface. This low pressure on the upper airfoil surface in $Q_2$ causes a positive pitch-up aerodynamic moment ($C_M>0$). Since this occurs during the pitch-up motion $\dot{\theta}>0$, we obtain a positive $\mathbf{W}_{Q_2}$ during the pitch-up phase. During the pitch-down phase (t/T=4.25) (figure~\ref{fig:CW_Q22}(i--iii)-(c),(g)), a similar low $C_p$-region also appears in $Q_2$. But since $\dot{\theta}<0$ in this phase, this pressure loading now contributes to a negative energy transfer to the foil. Hence, the low $C_p$ in $Q_2$ is responsible for a negative $\mathbf{W}_{Q_2}$ during the pitch-down motion and positive $\mathbf{W}_{Q_2}$ during the pitch-up motion as seen in figure~\ref{fig:CW_Q21}(a),(b). This is also reflected as negative $\mathbf{w}_1$ on the upper surface shear layer during the pitch-down phase (see zoomed-in inset in figure~\ref{fig:CW_Q22}(ii)-(a)), and similarly a positive $\mathbf{w}_1$ on the upper surface shear layer during the pitch-up phase (see zoomed-in inset in figure~\ref{fig:CW_Q22}(ii)-(b)). 

For the airfoil-only system, $C_p$ near the upper surface leading edge is lower during the pitch-down phase ($t/T=4.25$, figure~\ref{fig:CW_Q22}(i--iii)(c)) compared to the pitch-up phase ($t/T=4.75$, figure~\ref{fig:CW_Q22}(i--iii)(d)). This is very evident for $M_\infty=0.8$ as we see a lower $C_p$ near the leading edge on the upper airfoil surface in $Q_2$ in figure~\ref{fig:CW_Q22}(iii)-(c) compared to figure~\ref{fig:CW_Q22}(iii)-(d). Since $C_p$ in $Q_2$ is lower during the pitch-down motion, the negative energy transfer to the foil in the pitch-down phase is greater than the positive energy transfer in the pitch-up phase thereby yielding an overall negative $E^\star_{Q_2}$ for all the Mach numbers for the airfoil-only system as shown in figure~\ref{fig:Rppm_hist}(c),(f). 

In contrast, for the cylinder-airfoil system, $C_p$ near the airfoil leading edge in $Q_2$ is lower during the pitch-up phase ($t/T=4.75$, figure~\ref{fig:CW_Q22}(i--iii)-(h)) compared to the pitch-down phase ($t/T=4.25$, figure~\ref{fig:CW_Q22}(i--iii)-(d)). As a result, unlike the airfoil-only system, the cylinder-airfoil system has a net positive $E^\star_{Q_2}$ (figure~\ref{fig:Rppm_hist}(d),(h)). This change in $Q_2$ is due to the blockage effects induced by the cylinder which results in flow acceleration on the airfoil upper surface. During the pitch-down motion $(t/T=4.25)$, the flow acceleration results in an increased flow separation on the upper surface for the cylinder-airfoil system (figure~\ref{fig:CW_Q22}(i--iii)-(e)) compared to the airfoil-only system (figure~\ref{fig:CW_Q22}(i--iii)-(a)). During the pitch-up motion $(t/T=4.25)$, the flow acceleration creates stronger upstream propagating shock/compression waves on the upper airfoil surface as indicated by the lower $C_p$ in $Q_2$, in the cylinder-airfoil system (figure~\ref{fig:CW_Q22}(i--iii)-(h)) compared to the airfoil-only system (figure~\ref{fig:CW_Q22}(i--iii)-(d)). This effect is especially evident for $M_\infty=0.6$. The airfoil-only system (figure~\ref{fig:CW_Q22}(i)) is entirely subsonic and and vortex formation on the airfoil surface is absent. On the other hand, the cylinder-airfoil system develops weak compression waves on the airfoil upper surface during the pitch-up motion (indicated by an arrow in figure~\ref{fig:CW_Q22}(i)-(f,h)). Thus, the cylinder-induced blockage effects, causes flow acceleration on the upper airfoil surface, results in a lower the pressure region during the pitch-up motion than pitch-down motion, resulting in $E^\star_{Q_2}>0$.

In $Q_3$, the pitch-down phase is the energy-extraction phase for both systems (see figure~\ref{fig:CW_Q21}(d),(e)). For both systems, a low $C_p$ region near the airfoil leading edge in $Q_3$ produces a negative aerodynamic moment, $C_M<0$. This negative $C_M$ during the pitch-down motion $\dot{\theta}<0$ results in a positive energy transfer to the foil, $\mathbf{W}_{Q_3}>0$. Conversely, during the pitch-up motion when $\dot{\theta}>0$, a similar negative aerodynamic moment results in negative energy transfer to the foil, $\mathbf{W}_{Q_3}<0$. For the airfoil-only system, $C_p$ is lower near the airfoil leading edge in $Q_3$ during the pitch-up phase ($t/T=4.75$, figure~\ref{fig:CW_Q22}(i--iii)-(d)) than during the corresponding pitch-down phase ($t/T=4.25$, figure~\ref{fig:CW_Q22}(i--iii)-(c)). This lower $C_p$ in $Q_3$ results in a larger negative energy transfer during the pitch-up phase than the positive energy transfer during the pitch-down phase. Hence, the airfoil-only system shows $E^\star_{Q_3}<0$ at all $M_\infty$ as shown in figure~\ref{fig:Rppm_hist}(c),(f). 

For the cylinder-airfoil system (figure~\ref{fig:CW_Q22}(i--iii)-(g,h)), the addition of the cylinder results in a lower $C_p$ in $Q_3$ compared to the airfoil-only case (figure~\ref{fig:CW_Q22}(i--iii)-(c,d)). The cylinder shear and the wake introduce additional low-pressure regions as seen in $C_p$ fields (figure~\ref{fig:CW_Q22}(i--iii)-(g,h)). Additionally, the gap region between the cylinder and the airfoil lower surface, causes the flow acceleration, also resulting in a lower $C_p$ underneath the airfoil. 

Additionally, for the cylinder-airfoil system, we observe that $C_p$ on the airfoil lower surface in $Q_3$ is lower during the pitch-down phase ($t/T=4.25$, figure~\ref{fig:CW_Q22}(i--iii)-(g)) than during the pitch-up phase ($t/T=4.75$, figure~\ref{fig:CW_Q22}(i--iii)-(h)). This difference is most evident at $M_\infty=0.8$ (figure~\ref{fig:CW_Q22}(iii)-(g,h)). At $M_\infty=0.8$, the cylinder wake shedding is suppressed and the wake region is elongated. The interaction region between the airfoil lower-surface shear layer and the elongated cylinder wake introduces a shock-train-like region. During the pitch-down phase, at $t/T=4.25$ (figure~\ref{fig:CW_Q22}(iii)-(e,g)), this interaction region is associated with predominantly positive $\mathbf{w}_1$ and a low $C_p$. During the pitch-up phase ($t/T=4.75$, figure~\ref{fig:CW_Q22}(iii)-(h)), the shock-train-like region is still present in $Q_3$ but the $C_p$ on the airfoil lower surface is higher than during the pitch-down phase (figure~\ref{fig:CW_Q22}(iii)-(g)). 

The differences in the pressure loading on the airfoil surface in $Q_3$ arise due to the asymmetry in interaction region for the cylinder-airfoil system. During the pitch-down motion, the incidence angle in the gap region results in a higher flow acceleration than during the pitch-up motion. Hence the $C_p$ in the gap region in lower at $t/T=4.25$ is lower then $t/T=4.75$. This asymmetry in $Q_3$ is also seen for $M_\infty=0.6$ and 0.7 (figure~\ref{fig:CW_Q22}(i--ii)-(g,h)). Since the cylinder recirculation region length is reduced as $M_\infty$ reduces, the gap region length also reduces for $M_\infty=0.6$ and 0.7 when compared with $M_\infty=0.8$ (also see figure~\ref{fig:AoA5_flow}(g)-(l)). At these $M_\infty$, we also observe flow separation and shedding on the airfoil lower surface in $Q_3$. For $M_\infty$ = 0.6 and 0.7, this shortened gap region coupled with flow separation also results in a positive $C_p$ on the airfoil lower surface during the pitch-up phase (figure~\ref{fig:CW_Q22}(i--ii)-(h)). Therefore, the asymmetry in the gap-region results in a lower $C_p$ on the airfoil lower surface during the pitch-down motion, generating a net positive $E^\star_{Q_3}$ for the cylinder-airfoil system at all Mach numbers (figure~\ref{fig:Rppm_hist}(d),(g)).

\subsubsection{Phenomenology-based partitioning of $Q_3$ contributions}

As highlighted above, the $Q_3$ quadrant plays a dominant role in the energy transfer mechanisms of the cylinder--airfoil configuration. However, this quadrant encompasses several distinct flow phenomena, including the airfoil leading-edge shear layer, the cylinder shear layer, the wake--airfoil interaction region, and the shock-train structure. To isolate the individual contributions of these mechanisms to $E^\star_{Q_3}$, we perform a phenomenology-based sub-regional power partitioning of the $Q_3$ quadrant.
\begin{figure}
\centering
    \includegraphics[width=1.0\textwidth]{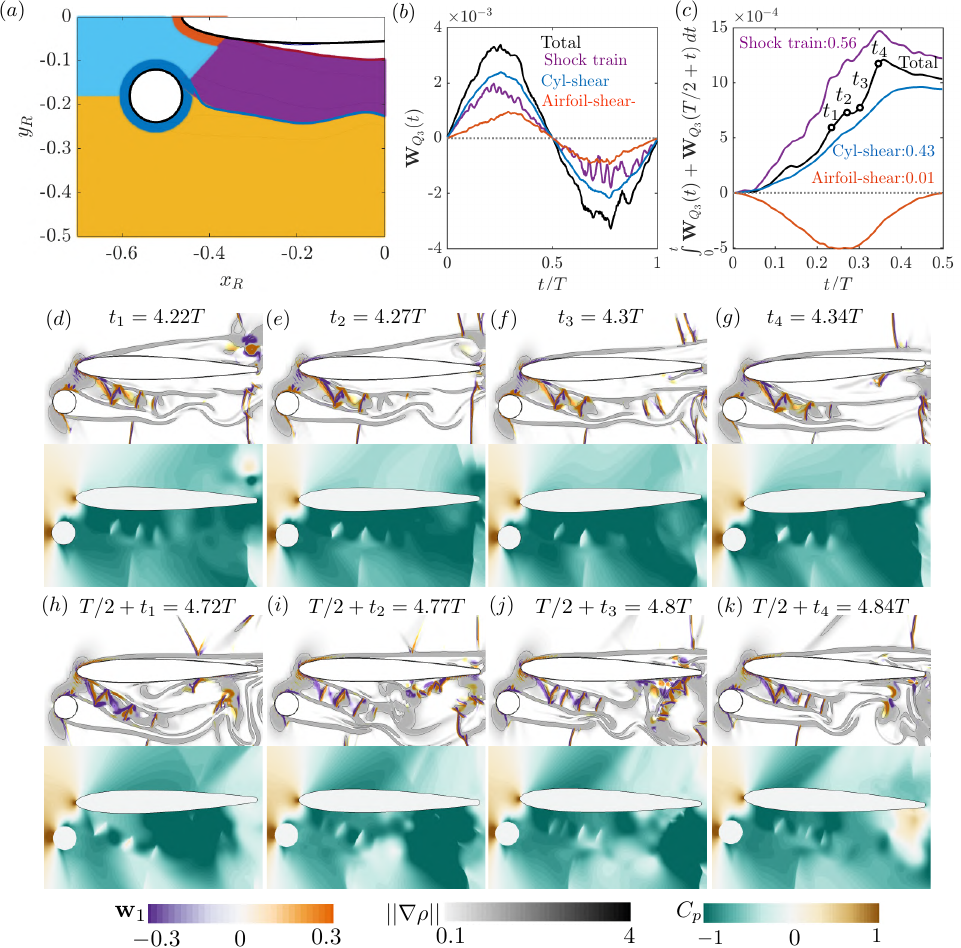}
    \caption{Sub-regional partitioning in $Q_3$ to isolate energy extraction contribution from distinct flow structures and phenomenologies. (a) Instantaneous sub-region division of $Q_3$ for a pitching cylinder-airfoil system at $(M_\infty,A_\theta)=(0.8,5^\circ)$. (b) Phase-averaged contribution from the shock-train, cylinder and airfoil shear layers to $\mathbf{W}_{Q_3}$. (c) Cumulative integral of the asymmetric component $\mathbf{W}_{Q_3}$ for the subregions. (d)-(k) Power density fields visualized using $\mathbf{w}_1$, $||\nabla \rho||$ in gray and $||\nabla \rho||=3$ contour line, and $C_p$-fields at highlighted time instances $t_i$ and $T/2+t_i$ in (c). }
    \label{fig:CW_Q3_sub}
\end{figure}

Figure~\ref{fig:CW_Q21}(e) shows that the cylinder--airfoil system exhibits qualitatively similar trends in $\mathbf{W}_{Q_3}$ across all $M_\infty$. In particular, the $M_\infty=0.8$ case is especially useful because the cylinder wake shedding is largely suppressed, allowing the interaction-region dynamics to be isolated without the additional complexity introduced by large-scale wake vortices. We therefore employ this case to define the sub-regional decomposition of $Q_3$ shown in figure~\ref{fig:CW_Q3_sub}(a).

The $Q_3$ quadrant is divided into five sub-regions associated with the dominant local flow features: the leading-edge airfoil shear layer (orange), the cylinder shear region (blue), the shock-train region (purple), the gap-flow region upstream of the shock train (cyan), and the region beneath the cylinder (yellow). The cylinder shear region and leading-edge shear region are identified by extruding the corresponding solid surfaces with an offset distance of $0.025c$ along the local surface-normal direction. The upper and lower boundaries of the shock-train region are defined by the airfoil lower-surface shear layer and the upper cylinder shear layer, respectively, where these shear layers are identified using the vorticity contours $w_z=\pm10$. The left boundary of the shock-train region is determined from the intersection of the $||\nabla \rho||=3$ contour with the identified cylinder and airfoil shear regions, while the right boundary coincides with the edge of the $Q_3$ quadrant at $x=0$.

 The phase-averaged contributions from these sub-regions to $\mathbf{W}_{Q_3}$ are shown in figure~\ref{fig:CW_Q3_sub}(b). We note that among the five sub-regions, the shock-train, cylinder-shear and the airfoil-shear regions contribute to positively to $E^\star_{Q_3}$. The cylinder-shear region exhibits the largest variation through the cycle followed by the shock-train and the airfoil-shear regions. Since we are interested in the net energy transfer to the airfoil for the pitching motion, we consider the cumulative integral of the asymmetric component of $\mathbf{W}_{Q_3}$ in figure~\ref{fig:CW_Q22}(c). This asymmetric component is obtained by adding ${\mathbf{W}}_{Q_3}$ at a pitch-down instant, $t$, and the corresponding pitch-up instant, ($T/2+t$). The cumulative integral of $\mathbf{W}_{Q_3}$ over the total $Q_3$ region (black line in figure~\ref{fig:CW_Q3_sub}(c)) converges to a lower value than the shock-train region (purple). This occurs because the other sub-regions corresponding to the upstream gap and the region underneath the cylinder contribute negatively to $\mathbf{W}_{Q_3}$. The shock-train region is the dominant source of positive energy transfer in $Q_3$, accounting for nearly 56\% of the total positive contribution. Furthermore, despite its greater distance from the airfoil surface, the cylinder shear-layer region provides the second-largest contribution at 43\%, while the leading-edge shear region of the airfoil provides a surprisingly small contribution to the positive energy transfer in $Q_3$. This phenomenology-based partitioning therefore reveals several counterintuitive features of the energy-transfer process: despite being closest to the airfoil surface, the leading-edge shear layer contributes only weakly to the positive power input, whereas the shock-train region and the more remote cylinder shear-layer structures emerge as the dominant contributors, thereby highlighting the unexpectedly strong influence of compressible gap-flow interactions on the flutter dynamics.

We consider a few time instances during the pitch-down and the corresponding pitch-up motion to further analyze the asymmetry. The power density fields $\mathbf{w}_1$, and the corresponding $C_p$ fields at four time instances within the time range from $t/T\in[4.2,4.35]$ are shown in figure~\ref{fig:CW_Q3_sub}(d)-(g). The main difference between the pitch-down (figure~\ref{fig:CW_Q3_sub}(d)-(g)) and the pitch-up (figure~\ref{fig:CW_Q3_sub}(h)-(k)) phases in $Q_3$ is the structure of the shock-train-like region and the flow separation on the airfoil lower surface. During the pitch-down motion (figure~\ref{fig:CW_Q3_sub}(d)-(g)), the cylinder shear layer region and the shock-train region predominantly result in positive $\mathbf{w}_1(\boldsymbol{x},t)$ and a low $C_p$ region on the airfoil lower-surface. The corresponding pitch-up instances, $t/T\in[4.65,4.8]$ (figure~\ref{fig:CW_Q3_sub}(h)-(k)), show a significant flow separation and shedding near the trailing edge during the pitch-up motion. During the pitch-up phase, at $t/T=4.72$ (figure~\ref{fig:CW_Q3_sub}(h)), flow separation on the airfoil lower surface starts from the leading edge and we observe a vortex roll-up near the trailing edge. This flow separation then results in vortex shedding near the trailing edge at $t/T=4.77,4.8$ (figure~\ref{fig:CW_Q3_sub}(i,j)). During this flow separation and shedding process on the airfoil lower surface (figure~\ref{fig:CW_Q3_sub}(h)-(j)), the $C_p$ in $Q_3$ is higher near the airfoil leading edge than during the pitch-down motion (figure~\ref{fig:CW_Q3_sub}(d)-(g)). The asymmetry in the airfoil lower-surface flow induced by the cylinder between the pitch-up and pitch-down motions modifies the shock-train structure and the pressure loading, producing a net positive energy contribution from $Q_3$. Overall, the sub-regional partitioning identifies the shock-train interaction region as the primary mechanism responsible for the positive net energy transfer ($E^\star_{Q_3}>0$).

\subsection{Influence of cylinder placement}

\begin{figure}
\centering
    \includegraphics[width=0.95\textwidth]{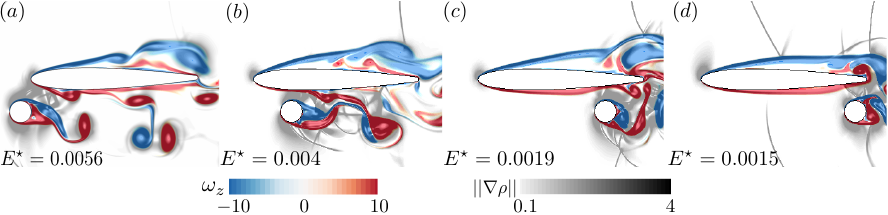}
    \caption{Effect of underwing cylinder placement on energy extraction from fluid to the airfoil. (a)-(d) Instantaneous vorticity-fields of pitching cylinder-airfoil systems with different horizontal cylinder placements of $x_{c0}=\{-0.522,-0.25,0.25,0.4\}$ and $y_{c0}=-0.18$ corresponding to $(M_\infty,A_\theta)=(0.7,5^\circ)$ at $t/T=0.25$. Flowfields are visualized using $\omega_z$ and $||\nabla \rho||$. }
    \label{fig:flow_cyl}
\end{figure}

So far we consider a cylinder placed upstream of the pivot point. We now consider various horizontal placements of the cylinder relative to the pivot point on the airfoil to examine the effect of that parameter. The instantaneous vorticity-fields at $\theta=0^\circ$ during the pitch-down motion are visualized in figure~\ref{fig:flow_cyl} for $(M_\infty,A_\theta)=(0.7,5^\circ)$. We observe that as the cylinder is moved downstream of the pivot point, the resulting $E^\star$ is significantly reduced. This indicates that the downstream cylinder placement shrinks the flutter boundaries. 

As the cylinder is moved downstream in the streamwise direction, the blockage effects from the cylinder that resulted in flow acceleration on the upper surface are reduced. This is evident as the flow separation region is moved towards the trailing edge as $x_{c0}$ is varied from -0.522 to 0.40 in figure~\ref{fig:flow_cyl}. Further, when the underwing cylinder is placed downstream of the pivot point, the shock-train and the gap region interact with the trailing edge vortex shedding. All these effects result in a significantly smaller flutter boundary as the underwing cylinder is placed downstream of the airfoil mid-chord. 

\section{Conclusions}
\label{sec:conclusions}

We investigated the aerodynamic effects of wake--wing interaction on transonic flutter using high-fidelity simulations of a sinusoidally pitching NACA0012 airfoil with and without an underwing cylinder at a Reynolds number of 10,000. The simulations were performed using a sharp-interface immersed boundary method over a range of transonic Mach numbers.

Energy-map analysis showed that the addition of the underwing cylinder significantly broadens the flutter boundary relative to the airfoil-only configuration. Whereas flutter in the airfoil-only system is primarily associated with shock-induced trailing-edge separation, the cylinder--airfoil system exhibits substantially more complex flow interactions involving wake-induced acceleration, gap-flow dynamics, and shock-train formation.

A major contribution of this work is the extension of force and power partitioning frameworks to compressible flows with moving boundaries. The compressible power-partitioning framework enabled direct identification of the flow structures and mechanisms responsible for net energy transfer to the wing. The dominant contribution was found to arise from the volumetric Lagrangian rate of change of momentum term in the conservation of momentum equation. Through quadrant-wise and sub-regional partitioning, we showed that the cylinder fundamentally alters the flutter mechanism by introducing two coupled effects. First, blockage induced by the upstream cylinder accelerates the flow over the upper airfoil surface, enhancing energy extraction during the pitch-up phase. Second, the gap between the cylinder shear layer and the lower-surface leading-edge shear layer forms a confined acceleration region that generates a shock train. Among the identified structures, the shock-train region is the dominant source of energy extraction, followed by the cylinder shear layer. In contrast to the airfoil-only system, the cylinder wake also suppresses lower-surface shock-induced separation.

The study further showed that cylinder placement strongly influences flutter behavior. An upstream cylinder enhances instability and energy transfer to the wing, whereas placing the cylinder downstream of the pivot point substantially reduces the extracted energy. These findings demonstrate that upstream wake disturbances can strongly intensify transonic flutter through coupled vortex, shear-layer, and compressibility effects.

More broadly, the compressible force- and power-partitioning framework developed here provides a new physics-based tool for analyzing unsteady high-speed flows involving fluid--structure interaction. Beyond the present flutter problem, the methodology has potential applications in shock-boundary-layer interactions, turbomachinery flows, aeroelasticity, rotorcraft aerodynamics, propulsion systems, and other compressible flow configurations where coherent structures govern unsteady loading and energy transfer.

\section*{Acknowledgments}
	\label{sec:acknowledgments}
The authors acknowledge the support from the U.S. Air Force Office of Scientific Research (Grant number: FA9550-23-1-0010). We also acknowledge the U.S. Department of Defense High Performance Computing Modernization Program (HPCMP) for computational resources.
	
	\section*{Declaration of interest}
	\label{sec:doi}
	The authors report no conflict of interest.

 \appendix

\section{Grid and statistical convergence}\label{appA}

\begin{figure}
\centering
    \includegraphics[width=1.0\textwidth]{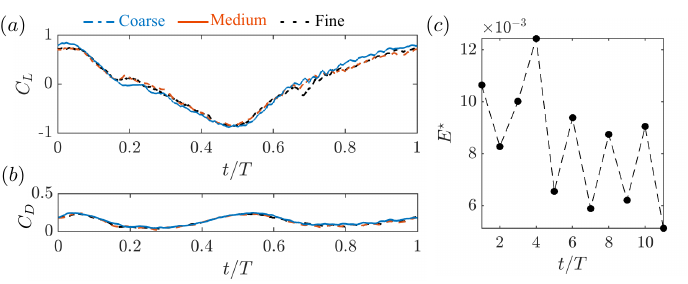}
    \caption{Grid and statistical convergence. (a)-(b) Grid dependence test for pitching cylinder-airfoil system at $(M_\infty,A_\theta)=(0.7,10^\circ)$ using phase-averaged $C_L$ and $C_D$ of the airfoil. (c) Variation of $E^\star$ for each pitching cycle.}
    \label{fig:grid}
\end{figure}

The grid verification test is conducted for the pitching cylinder-airfoil system at $Re=10,000$, $(M_\infty,A_\theta)=(0.7,10^
\circ)$. The details of the considered grids are presented in table~\ref{tab:grid}. We consider the phase-averaged lift and drag coefficients, $C_L$ and $C_D$, of the airfoil over 5 pitching cycles (figure~\ref{fig:grid}(a), (b)). We observe RMS errors of 2.6\% and 0.7\% for the phase-averaged lift and drag coefficients between the medium and fine grid. We hence use the medium grid for the current study.

To examine the statistical convergence of the simulations, the cycle-to-cycle variation of $E^\star$ and the five cycle moving average of $E^\star$ for the pitching cylinder-airfoil system at $(M_\infty,A_\theta)=(0.7,10^\circ)$ are shown in figure~\ref{fig:grid}(c). We consider the first four pitching cycles as transient and base  all the analysis in the paper on the subsequent five cycles.
\begin{table}
  \centering
  \caption{Grid resolutions considered in the present study.}
  \label{tab:grid}
  \begin{tabular}{lccl}
    \toprule
    Grid & $N_x$ & $N_y$ & $\Delta x_{\min}$ \\
    \midrule
    Coarse & 800  & 520  & 0.0025   \\
    Medium & 1240 & 960  & 0.00125  \\
    Fine & 1808 & 1454 & 0.0008 \\
  \end{tabular}
  \label{tab:grid}
\end{table}

\section{Effects Associated with Quadrants $Q_1$ and $Q_4$}\label{appB}
In the main paper, we have focused on $Q_2$ and $Q_3$ which dominate the enhancement of energy transfer in the cylinder-airfoil system (see figure~\ref{fig:Rppm_hist}).  For completeness sake, we present the effects associated with $Q_1$ and $Q_4$, which provide smaller but non-negligible contributions to the energy transfer.
\begin{figure}
\centering
    \includegraphics[width=0.95\textwidth]{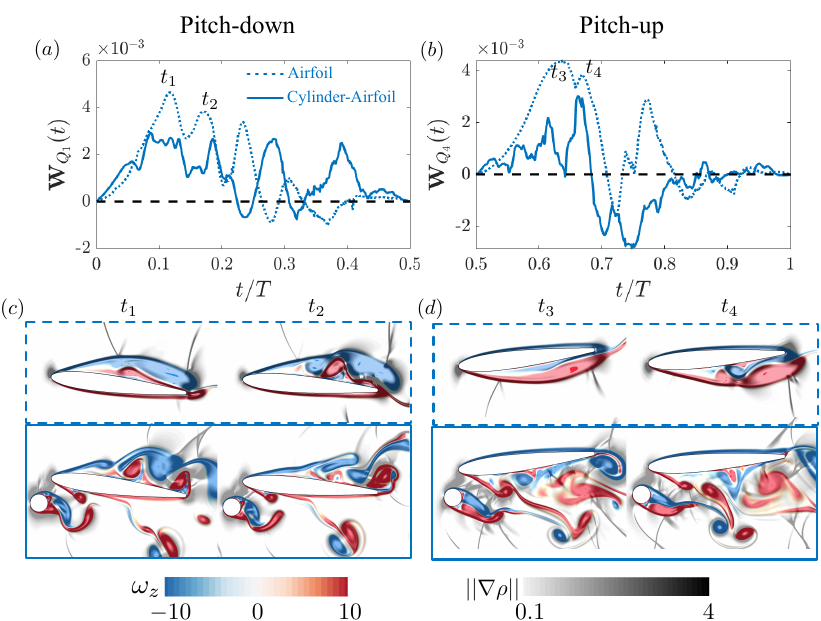}
    \caption{Effect of underwing cylinder on shock-induced separation in $Q_1$ and $Q_4$. (a)-(b) Phase-averaged $\mathbf{W}_{Q_1}$ and $\mathbf{W}_{Q_4}$ for airfoil and cylinder-airfoil system during the pitch-down and pitch-up motions, respectively at $(M_\infty,A_\theta)=(0.7,10^\circ)$ (c)-(d) Instantaneous vorticity-fields visualized using $\omega_z$ and $||\nabla \rho||$ at highlighted time instances. }
    \label{fig:CW_Q1Q2}
\end{figure}

For the flow over pitching airfoil-only system, shock-induced separation is the primary flutter instability mechanism and is seen in the positive $E^\star_{Q_1}, E^\star_{Q_4}$ at $M_\infty=0.7$ and both $A_\theta=5^\circ$ and $10^\circ$ (figure~\ref{fig:Rppm_hist}(c),(f)). The addition of an upstream underwing cylinder significantly reduces this energy transfer to the foil in $Q_4$ (figure~\ref{fig:Rppm_hist}(e),(h)) as seen as the negative $\Delta E^\star_{Q_4}$. This shows that the upstream underwing cylinder wake suppresses the shock-induced stall mechanism in $Q_4$. In $Q_1$, $\Delta E^\star_{Q_1}$ is negative for $A_\theta=10^\circ$ and negligible at $A_\theta=5^\circ$ indicating a more subtle effect on the shock-induced stall mechanism in $Q_1$.  

The phase-averaged $\mathbf{W}_{Q_1}$ and $\mathbf{W}_{Q_4}$ for the cylinder-airfoil system and the airfoil are plotted during the pitch-down and pitch-up motions respectively in figure~\ref{fig:CW_Q1Q2}(a)-(b). From the vorticity-fields (figure~\ref{fig:CW_Q1Q2}(c)-(d)), the flow separation occurs mainly near the trailing edge region for the airfoil-only system during the pitch-down $(t_1,t_2)$ and the pitch-up $(t_3,t_4)$ motions. For the cylinder-airfoil system, in $Q_1$ the flow separation occurs more upstream near the leading edge than the airfoil-only system (figure~\ref{fig:CW_Q1Q2}(c)). Additionally, several weaker vortices are shed when the upstream cylinder is present whereas we observe a more coherent roll-up and shedding in the airfoil-only system. This results in a decreased energy contribution from the shock-induced separation on the upper airfoil surface.

A significant difference due to the cylinder presence is observed in $Q_4$ (figure~\ref{fig:Rppm_hist}(e),(h)). Since the shock-induced separation for the airfoil-only system occurs during the pitch-down motion in $Q_4$, the vorticity-fields during the pitch-down motion at $t_3$ and $t_4$ are depicted in figure~\ref{fig:CW_Q1Q2}(d). The cylinder presence directly impacts the formation on the $\lambda-$shock wave system on the lower airfoil surface. The cylinder wake suppresses the shock-induced flow separation seen in the airfoil-only case, thereby resulting in a negative net energy transfer in $Q_4$ (figure~\ref{fig:Rppm_hist}(g)) for $M_\infty=0.7,A_\theta=10^\circ$.

%\section{Compressible force and power partitioning method}\label{appB}
 \bibliographystyle{unsrtnat}
 \bibliography{refs}

\end{document}